\documentclass{osa-article}

\journal{osajournal}
\articletype{Research Article}
\usepackage{color}
\newcommand{\edit}[1]{{\color{black}{#1}}}    % Uncomment if you want to see editing as red
\begin{document}

\title{Influence of disorder on a Bragg microcavity}

\author{S. G. Tikhodeev,\authormark{1,2,*} E. A. Muljarov,\authormark{3} W. Langbein,\authormark{3}
N. A. Gippius,\authormark{4} H. Giessen,\authormark{5} and T. Weiss,\authormark{5}}

\address{\authormark{1}M. V. Lomonosov Moscow State University, Leninskie Gory 1, Moscow 119991, Russia\\
\authormark{2}A. M. Prokhorov General Physics Institute, Russian Academy of Sciences, Vavilova Street 38, Moscow 119991, Russia\\
\authormark{3}Cardiff University, School of Physics and Astronomy, The Parade, CF24 3AA, Cardiff, United Kingdom\\
\authormark{4}Skolkovo Institute of Science and Technology, Nobel Street 3, Moscow 143025, Russia\\
\authormark{5}4th Physics Institute and Research Center SCoPE, University of Stuttgart, Stuttgart D-70550, Germany}

\email{\authormark{*}tikh@gpi.ru} %% email address is required

 \homepage{https://doi.org/10.1364/JOSAB.402986.\\
 \edit{ \rm{Journal of the Optical
Society of America B, Vol. 38, No. 1, pp. 139--150 (2021)}}} %% author's URL, if desired


\begin{abstract}
Using the resonant-state expansion for leaky optical modes
of a planar Bragg microcavity, we investigate the influence of disorder
on its fundamental cavity mode. We model the disorder by randomly
varying the thickness of the Bragg-pair slabs (composing the mirrors) and the cavity,
and calculate the resonant energy and linewidth of each disordered microcavity
exactly, comparing the results with the resonant-state expansion for a
large basis set and within its first and second orders of perturbation theory.
We show that random shifts of interfaces
cause a growth of the inhomogeneous broadening of the
fundamental mode that is proportional to the
magnitude of disorder. Simultaneously, the quality factor of the microcavity
decreases inversely proportional to the square of the magnitude of disorder.
We also find  that first-order perturbation theory works very
accurately up to a reasonably large disorder magnitude, especially for calculating the
resonance energy, which allows us to derive qualitatively the scaling  of
the microcavity properties with disorder strength.
\end{abstract}

%%%%%%%%%%%%%%%%%%%%%%%%%%  body  %%%%%%%%%%%%%%%%%%%%%%%%%%

\section{\label{Sec1} Introduction}

Disorder plays an important role in photonics.  For example, it drives the coloring and polarization conversion of natural  disordered
light diffusers such as opals, birds feathers, or wings of butterflies~\cite{Vukusic2003,Zi2003,Kinoshita2008,Wiersma2013,Wu2018}.
Unavoidable technological imperfections can sometimes critically reduce the
desired performance of photonic crystal slab
waveguides and nanocavities~\cite{Gerace2005,Taguchi2011,Ashida2018,Mohamed2018}. Different theoretical approaches have been proposed to
describe the role of disorder, either numerically~\cite{Demesy2007,Hagino2009} or based on various versions of
perturbation
theory in electrodynamics~\cite{Johnson2002,Gerace2005,Wiersig2017,Vasco2018,Vasco2018a}. The important prerequisite for any perturbation theory
is a suitable basis, which, in the case of open electrodynamical systems, is composed of resonant states (also known as quasi-normal or leaky
modes)~\cite{Leung1994,Tikhodeev2002,Gippius2005,Muljarov2010,Doost2012,Muljarov2016,Alpeggiani2017,Lassalle2018,Yan2018,Lalanne2018,Lalanne2019,Defrance2020}
that determine the resonant optical response,
e.g., the Fano resonances in  open cavities~\cite{Fano1941,Fano1961,Luk'yanchuk2010}.

Recently, the resonant-state expansion, a rigorous perturbation theory for calculating the resonant states  of
any open system in electrodynamics based on a finite number of resonant states of some more elementary system, has been developed~\cite{Muljarov2010}.
Originally proposed for purely dielectric  shapes (slabs, microspheres, microcavities~\cite{Doost2012}) with nondispersive dielectric permittivity,
the method was then generalized to
dispersive open systems~\cite{Muljarov2016a}, photonic crystal slabs~\cite{Neale2020}, and periodic arrays of nanoantennas at
normal~\cite{Weiss2016} and oblique incidence~\cite{Weiss2017},
and open systems containing
magnetic, chiral, or bi-anisotropic materials~\cite{Muljarov2018}.
In addition, the method has been extended to waveguide geometries such as dielectric slab waveguides~\cite{Armitage2014,Armitage2018}
and optical fibers~\cite{Upendar2018},
with a possibility to account for nonuniformities~\cite{Lobanov2017a} and nonlinearities~\cite{Allayarov2018,Allayarov2020}.

The perturbation in the
resonant-state expansion can be of any shape within the basis volume.
The difference from the basis reference can even be  huge
when using a sufficiently large number of resonant states as basis. In order to have
a  meaningful physical picture, it is, however, better to describe the structure of interest using a minimum number
of resonant states, see, e.g., examples of calculating the sensor performance with a single resonant state first-order approximation~\cite{Weiss2016,Both2019},
and the interaction of
spatially separated photonic crystal slabs with a pair of quasi-degenerate states in Ref.~\cite{Weiss2017}.

While full-wave simulations have been already used to investigate the resonant states in disordered media~\cite{Lalanne2018},
we concentrate in this paper on the impact of disorder on the resonant states of a Bragg microcavity
by comparing
 full-wave calculations with the resonant-state expansion as well as its
first- and second-order perturbative
formulations. In particular, we vary  randomly the thickness
of the Bragg-pair slabs (acting as the mirrors) and the cavity
itself, and derive how the resonant states change with growing amplitude of random displacements.
On the one hand, because of the simplicity of the system,
its disorder-modified states (their energies,
linewidths, and field distributions) can be calculated with any accuracy via
linearization of the frequency dependence of the inverse scattering matrix around the resonant state of
interest~\cite{Gippius2005,Gippius2010,Weiss2011} for each disorder realization.
On the other hand, we can calculate the same resonances using the resonant-state expansion for an increasing number of resonant states in the basis,
 and
then compare them with the exact values. Repeating the calculations many times and retrieving the statistically
averaged results yields relevant information about the influence of disorder on the optical properties of
the Bragg microcavities. A similar approach has been used in Ref.~\cite{Upendar2018} where the impact of disorder on the
effective index of propagating modes in photonic crystal fibers has been investigated via the resonant-state expansion and
compared with full-wave simulations.

The paper is organized as follows: The model of the disordered Bragg microcavity is described in Sec.~\ref{Sec2},
the formulation of the resonant-state expansion is given in Sec.~\ref{Sec3}. Section~\ref{Sec4} summarizes the results
of the comparison between the exact solutions and those obtained by the resonant-state expansion using different orders of perturbation theory.
Special attention is paid to the disorder-induced inhomogeneous broadening in the ensemble of disordered cavities
(Subsec.~\ref{SubInhomo}) and the analysis of the influence of disorder magnitude on the statistically averaged resonance energies and
their homogeneous linewidths (Subsec.~\ref{SubVsA}). Section~\ref{Sec5} contains a discussion of the obtained results,
which are summarized in Sec.~\ref{Sec6}. Details of the linearization scheme of calculating the
poles of the scattering matrix
are given in Appendix~\ref{AppA}.
The accuracy of the different orders of perturbation theory based on the resonant-state expansion, depending on the
magnitude of disorder is discussed in Appendix~\ref{SubError}.

\section{\label{Sec2} Model}

\begin{figure}[h!]
\centering\includegraphics[width=7cm]{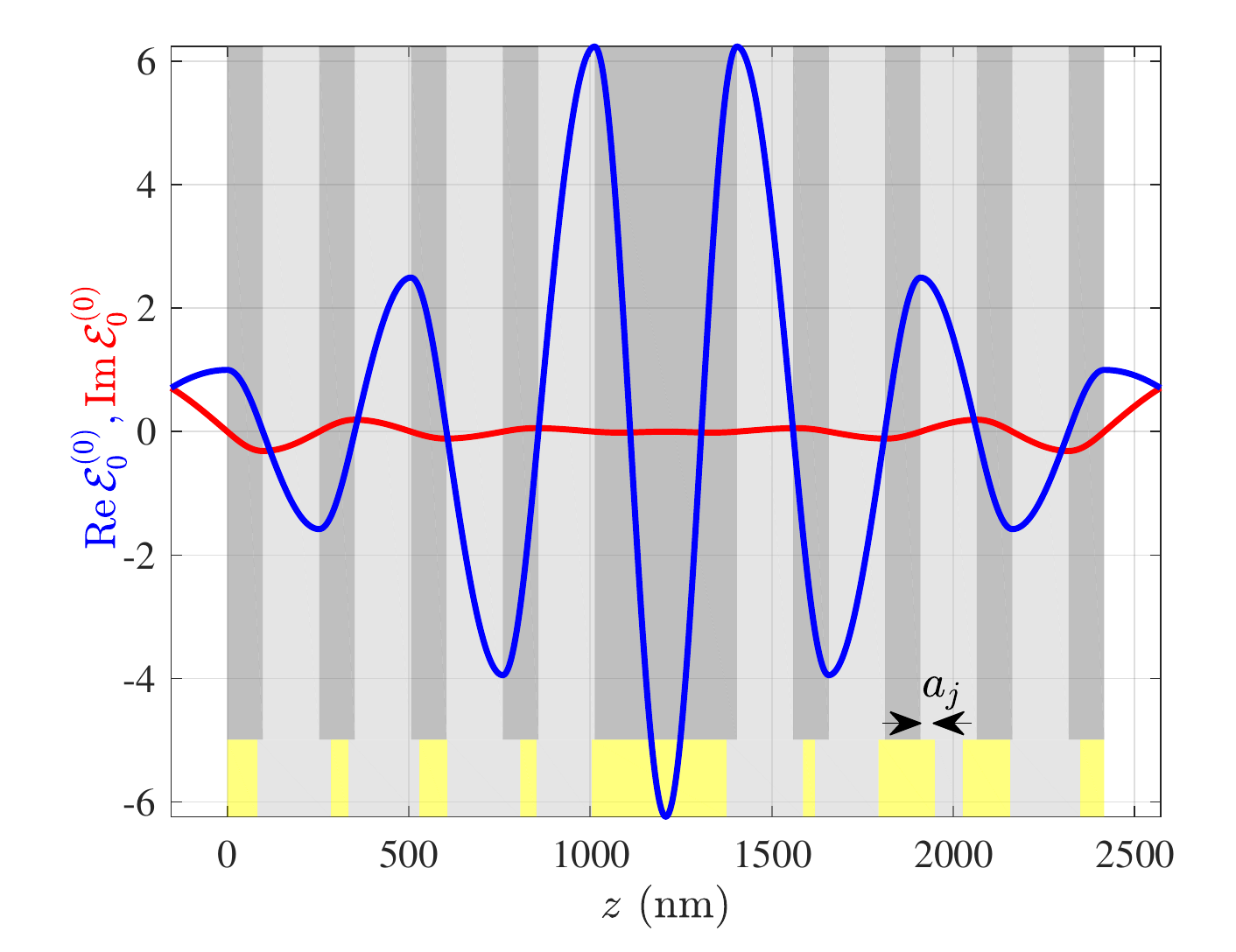}
\caption{\label{Fig1}
(Color online)
Schematic of the unperturbed Bragg microcavity (gray background) and spatial distributions of the real (blue solid line) and imaginary
(red solid line) parts of the
electric field $\mathcal{E}_0^{(0)}(z)$ of the fundamental cavity mode. Darker and brighter gray shades indicate
materials with dielectric susceptibilities $\varepsilon_1 = 10$
 and $\varepsilon_2 = 4$, respectively. Yellow/bright gray shades illustrate a realization of
 a microcavity with interfaces randomly
 displaced by shifts $a_j$, with disorder strength $a=0.5$  [see Eq.~(\ref{Eq_A})]
 }
\end{figure}

 \begin{figure}[h!]
\centering\includegraphics[width=7cm]{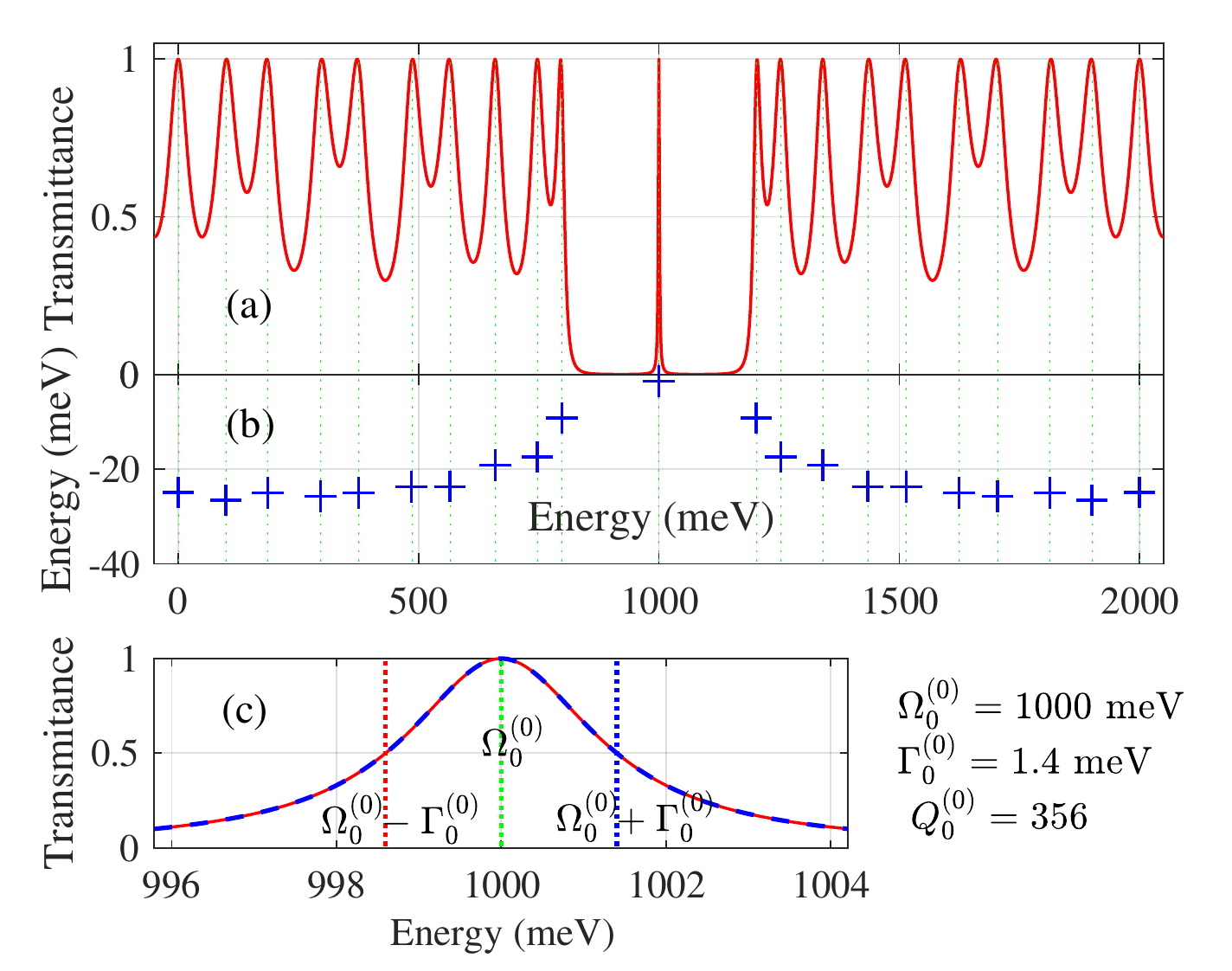}
\caption{\label{Fig2}
(Color online)
(a) Transmittance of the ideal microcavity as depicted in Fig.~\ref{Fig1}  (without random displacements).
(b) Map of resonant states of the microcavity on the complex energy plane (crosses); the vertical green dashed lines denote the positions
of the resonant states on the real energy axis.
(c) Transmission spectra in the vicinity of the fundamental resonance at 1~eV (red curve). The dashed blue curve shows the
single-pole resonant approximation given by Eq.~(\ref{T(E)}). Green, red, and blue vertical dashed lines mark
the energies $\Omega_0$, $\Omega_0-\Gamma_0$, and $\Omega_0+\Gamma_0$, respectively.}
\end{figure}

We consider a planar microcavity that is made of two Bragg mirrors
with $m$ pairs of layers of $\lambda/4$ optical thickness of nondispersive
materials with dielectric constants $\varepsilon_1$ and $\varepsilon_2$, surrounding a cavity layer of
 $M\times\lambda/2$ optical thickness of material with dielectric
constant $\varepsilon_1$. The cavity is surrounded by free space with permittivity $\varepsilon_0=1$.
In the numerical results presented we use $\varepsilon_1=10$,  $\varepsilon_2=4$,
$m=4$, and $M=2$, the latter corresponding to a  cavity layer of $\lambda$ optical thickness.
A schematic of the microcavity  is displayed in Fig.~\ref{Fig1}. We have chosen the parameters of the cavity such that
the fundamental cavity mode at normal incidence is $\Omega_0=2\pi{\hbar}c/\lambda=1$~eV ($\lambda = 1.24~\mu$m).
This corresponds to
thicknesses of the Bragg $\lambda/4$ layers of
$L_{1}=\pi c\hbar/(2\sqrt{\varepsilon_1}\Omega_0)\approx 98$~nm and $L_{2} =\pi c\hbar/(2\sqrt{\varepsilon_2}\Omega_0)\approx 155$~nm,
and the central cavity layer is $L_C=4L_1\approx 392$~nm thick.
Then the fundamental cavity mode
linewidth appears to be $2\Gamma_0 = 2.8$~meV corresponding to the quality factor $Q = \Omega_0/2\Gamma_0\approx 356$.
The spatial distributions $\mathrm{Re}\,\mathcal{E}_0$ and $\mathrm{Im}\,\mathcal{E}_0$  of
the resonant electric field of the fundamental cavity mode with eigenenergy $E_0 =\Omega_0-i\Gamma_0$
are shown in Fig.~\ref{Fig1}  by blue and red curves, respectively.

The optical scattering matrix of this simple microcavity (see in Appendix~\ref{AppA}) has an infinite series of discrete
Fabry-Perot poles on the complex energy plane, which manifest themselves as peaks in the transmission spectrum,
as shown in Figs.~\ref{Fig2}a,b.

The transmission spectrum in Fig.~\ref{Fig2}  has been calculated within a $2\times2$ optical scattering matrix approach as described
in Ref.~\cite{Tikhodeev2002} for homogeneous layers and normally incident light. More details are provided
in Appendix~\ref{AppA}. The poles
of the scattering matrix on the complex energy plane in Fig.~\ref{Fig2}b, as well as the
the electric eigenfields in Fig.~\ref{Fig1}  can be calculated via the
scattering matrix energy dispersion linearization\cite{Gippius2005,Gippius2010,Weiss2011}, as described in Appendix~\ref{AppA}.
The real part of the  eigenenergy, $\Omega_n = \mathrm{Re}\,E_n$, corresponds to the resonance energy,
while the imaginary part, $2\Gamma_n=-2\mathrm{Im}\,E_n$,
gives the resonance linewidth.
In what follows, we  mark the values corresponding to the
 unperturbed (ideal) microcavity by the upper index $(0)$, as shown
in Figs.~\ref{Fig1},\ref{Fig2}.

We now investigate the behavior of the
fundamental cavity mode denoted by eigenenergy $E_0$
under the influence of random displacements of the
microcavity interfaces. We will leave the external interfaces of the microcavity at their original positions, and assume that
all other $j = 1,2, \ldots J$ $(J=16)$ interfaces are shifted by
\begin{equation}\label{Eq_A}
a_j = a \beta_j L_1,
\end{equation}
where
%\begin{equation}\label{Eq_betas}
$\beta_1, \beta_2, \ldots \beta_{J} $ %= \mathrm{rnd}(-1, 1)
%\end{equation}
is a set of $J$ uniformly distributed uncorrelated random numbers within the interval (-1,1). The
disorder strength $a$ is chosen between zero and 0.5
in order to keep all resulting layer thicknesses positive.
Furthermore, we consider uncorrelated disorder with vanishing statistically averaged displacements
\begin{equation}\label{Eq_betaN}
\langle \beta_j \rangle = 0.
\end{equation}
Note that random shifts $a_j$ of the interfaces have been measured experimentally
before in disordered GaAs/AlAs cavities~\cite{Zajac2012}.

\section{\label{Sec3} Resonant-state expansion}

The resonant-state expansion~\cite{Muljarov2010,Doost2012,Doost2014,Weiss2017}  relies on knowing  the  electric
field distributions  $\mathcal{E}_n^{(0)}(z)$
of a set of resonant states with complex frequencies $E_n^{(0)}$ for a photonic structure with a spatial profile
of the dielectric susceptibility $\varepsilon^{(0)}(z)$. These resonant states are used in the resonant-state expansion as a basis to expand
the  electric fields of the resonant state of a modified structure with dielectric susceptibility
\begin{equation}\label{Eq_Eig1d}
\varepsilon(z)=\varepsilon^{(0)}(z)+\Delta\varepsilon(z)
\end{equation}
as
\begin{equation}\label{Eq_Eig1d}
\mathcal{E}(z)=\sum_n b_n \frac{\mathcal{E}_n^{(0)}(z)}{C_n}.
\end{equation}
The normalization constants $C_n$ have the analytical form~\cite{Muljarov2010,Weiss2017}
\begin{equation}\label{Eq_norm}
C_n^2 = \int_0^L \varepsilon (z) \mathcal{E}_{n}^2(z) dz +\frac{i}{2k_n}\left[\mathcal{E}_{n}^2(0)+\mathcal{E}_{n}^2(L) \right] ,
\end{equation}
where the range 0 to $L$ covers exactly the microcavity structure, and the fields in the second term have to be taken in the medium outside the
cavity. However, the fields are continuous at the outermost interfaces for the considered case of normal incidence, because this
results in purely transverse electric fields over the entire microcavity.
The general orthonormality of resonant states is given by~\cite{Muljarov2010,Weiss2017}
\begin{eqnarray}
\delta_{n'n} & = &\frac{1}{C_nC_n'} \left\{ \int_0^L \varepsilon (z) \mathcal{E}_{n'}(z)\mathcal{E}_n(z) dz\label{Eq_normnn} \right.\\ \nonumber
&& +\left.\frac{i}{k_{n'}+k_n}\left[\mathcal{E}_{n'}(0)\mathcal{E}_n(0)+\mathcal{E}_{n'}(L)\mathcal{E}_n(L)\right]\right\},
\end{eqnarray}
where $\hbar k_n = E_n/c$.

The coefficients $b_n$ and new eigenenergies $E$ can be calculated
via the linear eigenproblem~\cite{Muljarov2010}
\begin{equation}\label{Eq_beta}
\sum_{n'}W_{nn'}b_{n'} = E b_{n},
\end{equation}
where
\begin{equation}\label{Eq_Wn'n}
W_{nn'} = (A^{-1})_{nn'}E_{n'}^{(0)},
\end{equation}
\begin{equation}\label{Eq_An'n}
A_{nn'} = \delta_{nn'} + \frac{1}{2}V_{nn'},
\end{equation}
and the matrix elements of the perturbation are
\begin{equation}\label{Eq_Vnn'}
V_{nn'} = \frac{1}{C_nC_{n'}} \int_0^L \Delta \varepsilon (z) \mathcal{E}_{n}^{(0)}(z)\mathcal{E}_{n'}^{(0)}(z) dz.
\end{equation}
Following from the resonant-state expansion, the resonance eigenenergy in the first order of perturbation theory yields~\cite{Doost2014,Weiss2017}
\begin{equation}\label{Eq_En1st}
E_n^{(1)} \approx  E_n^{(0)}  \left( 1+\frac{1}{2}V_{nn}\right)^{-1},
\end{equation}
whereas  the  resonant state eigenenergy up to the second order of perturbation theory is given by~\cite{Doost2014}
\begin{equation}\label{Eq_En2nd}
E_n^{(2)}  \approx E_n^{(0)} \left( 1+\frac{1}{2}V_{nn}-\frac{1}{4}\sum_{n'\neq n}
     \frac{E_n^{(0)}V_{n'n}^2}{E_n^{(0)}-E_{n'}^{(0)}} \right)^{-1}.
\end{equation}

In the following, we keep explicitly the normalization constants $C_n$ in the resonant-state expansion formulas
and use the eigenfields (e.g., the one shown in Fig.~\ref{Fig1})
satisfying the conditions
\begin{equation}\label{Eq_rn}
 \mathcal{E}_n(0)=(-1)^{p_n}\mathcal{E}_n(L)=1 ,
\end{equation}
where $p_n =0,1$ denotes the eigenstate parity that is either even or odd due to the mirror symmetry of the
unperturbed cavity.
This choice of normalization is convenient for the calculation of the eigenfields within the
 linearization of the scattering matrix (see in Appendix~\ref{AppA}) and simplifies
the comparison with the resonant-state expansion.

\section{\label{Sec4} Influence of disorder}

While investigating the influence of disorder, we  compare the  scattering matrix result from linearization, which we call
here ``exact'', with the first- and second-order
approximations (\ref{Eq_En1st}) and (\ref{Eq_En2nd}) as well as with the full resonant-state expansion obtained by solving the eigenvalue
problem (\ref{Eq_beta}) with a truncation of
an infinite matrix. The resonant-state expansion is asymptotically exact, and its only limitation is the basis size.
The resonant states are calculated as described in Appendix~\ref{AppA}.
The basis size is taken as $N=419$ in the present paper, symmetrically around the fundamental cavity mode, see Appendix~\ref{AppA}.
  The same resonant states are used in the second-order perturbation theory.

  \begin{figure}[h!]
\centering{\includegraphics[width=7cm]{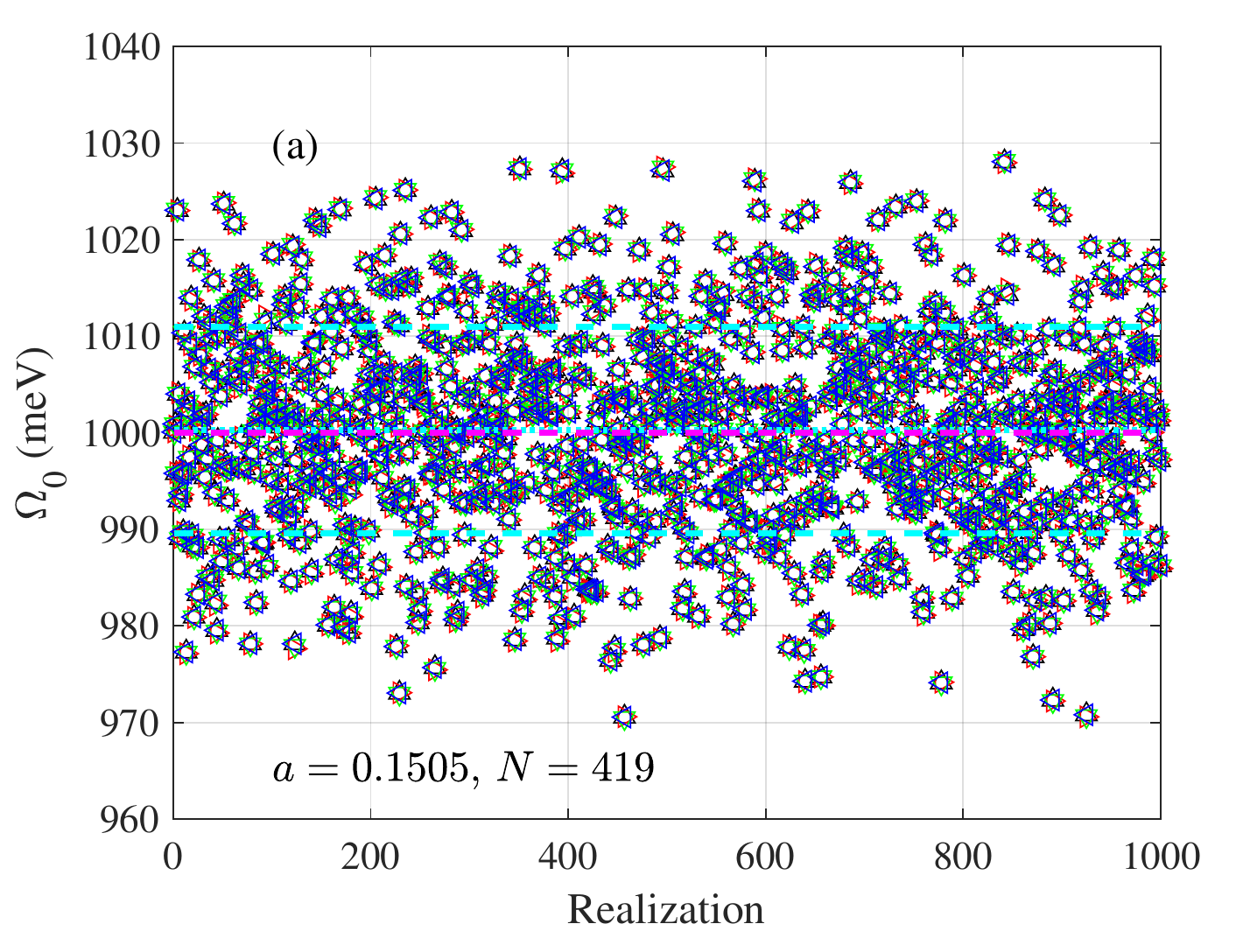}
\includegraphics[width=7cm]{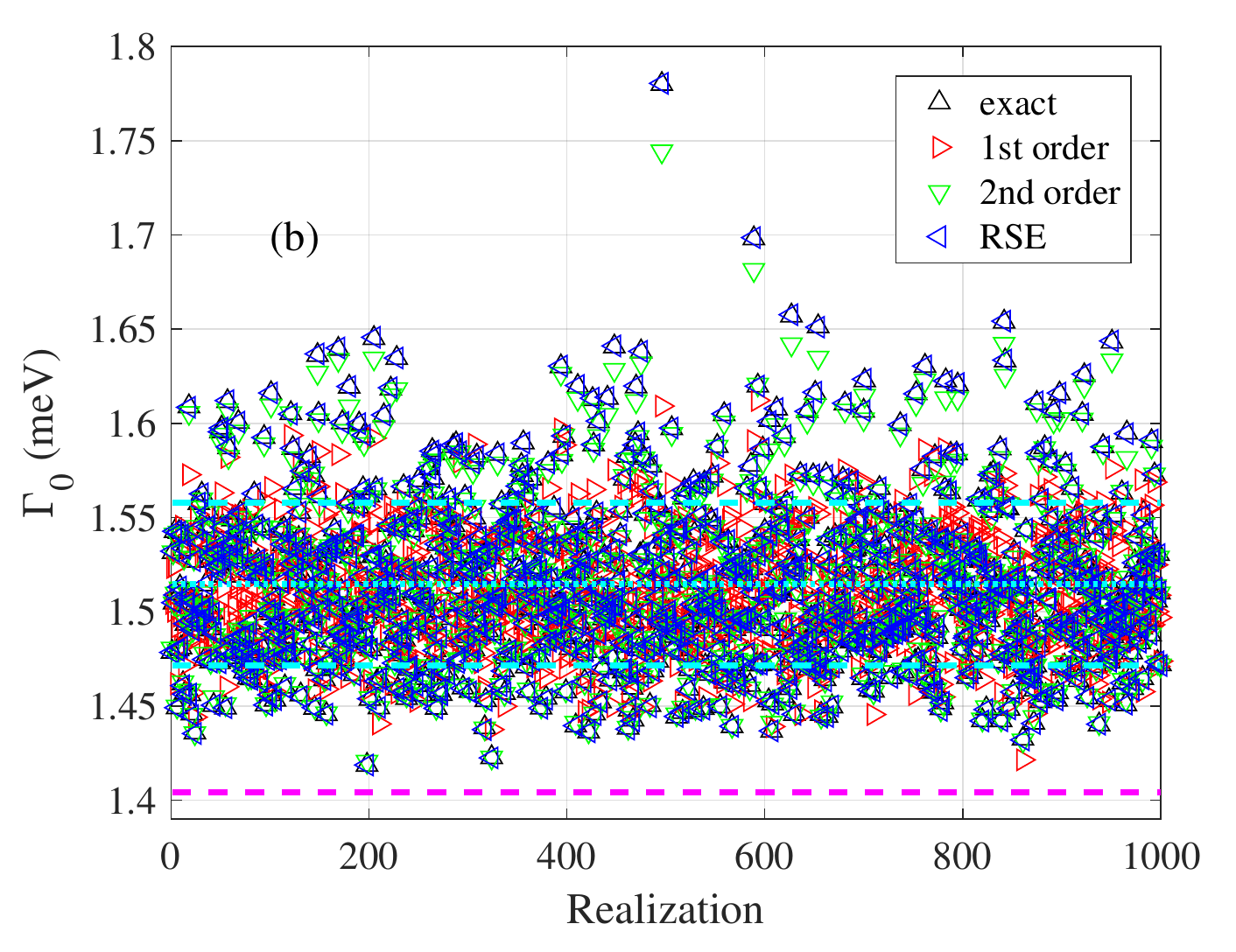}}
\caption{ \label{Fig3}
(Color online) Example of calculated resonance frequencies $\Omega_0$ (a) and half linewidths $\Gamma_0$ (b) of the fundamental microcavity resonance, for 1000 interface shift realizations
with disorder parameter $a=0.1505$.
The legend in panel (b) specifies the symbols for the different results: Exact calculation, first- and second-order perturbation theory [Eq.~(\ref{Eq_En1st})
and (\ref{Eq_En2nd}), respectively], and resonant-state expansion (RSE) [Eq.~(\ref{Eq_beta})] with 419 basis states.
Cyan dotted and dashed horizontal lines in panel (a) denote the
mean values $\langle \Omega_0\rangle$ and $\langle \Omega_0\rangle \pm \sigma_\Omega$, where $\sigma_\Omega$ is the standard deviation
of $\Omega_0$.
The magenta dashed horizontal line indicates the resonance energy $\Omega_0^{(0)}$ of the unperturbed microcavity.
Lines in panel (b) give the equivalent values for the resonance linewidth, i.e.,
$\langle\Gamma_0\rangle$, $\langle\Gamma_0\rangle\pm \sigma_\Gamma$, and $\Gamma_0^{(0)}$,
where $\sigma_\Gamma$ is the standard deviation of $\Gamma_0$.
 }
\end{figure}

Figure~\ref{Fig3}  illustrates changes of the real and imaginary parts of the fundamental cavity mode energy and linewidth for 1000 different
realizations of random shifts of interfaces with the disorder parameter $a=0.1505$. The latter means that the random displacements
of the interfaces are up to ${\sim}15$~nm.

It can be seen
that (\textit{i}) introducing disorder causes an inhomogeneous broadening of the resonance energy position, with a
standard deviation on the order of 10~meV; (\textit{ii}) the linewidth of the resonance (homogeneous broadening) grows
by approximately  10\% (from ${\sim}1.4$~meV to ${\sim}1.52$~meV); (\textit{iii}) the results for the resonance energy $\Omega_0$ (Fig.~\ref{Fig3}a),
calculated exactly, in the first and second perturbation orders, and in the resonant-state expansion do visually coincide for all disorder realizations,
while for the linewidth only the exact and the resonant-state expansion results coincide.

The difference, representing the calculation error between the exact results, the first, second and ``full'' resonant-state expansion (with $N=419$ resonant states in the basis)
is analyzed versus the disorder strength $a$ in Appendix~\ref{SubError}.
Since the absolute error is similar for the real and imaginary part, the relative error of calculating $\Omega_0$ is approximately $Q_0$ times smaller than that of $\Gamma_0$.

It is shown in Appendix~\ref{SubError} that the calculation error and its standard deviation
grow with $a$ and can be quite large,
especially for the first perturbation order.
Interestingly, the calculation errors for the quantities, averaged over many (1000 in this work) realizations,
remain relatively small over the investigated range of disorder parameter $a\leq 0.3$,
even in the first perturbation order.

In what follows we investigate the
statistics of $\Omega_0$ and $\Gamma_0$  as functions of the disorder parameter.
However, we begin from the analysis of the most visible effect of the disorder, namely
the inhomogeneous broadening of the fundamental cavity mode eigenenergy distribution due to the disorder.

\subsection{\label{SubInhomo} Inhomogeneous broadening}

The distribution of fundamental cavity mode energies $\Omega_0(\nu)$ (where $\nu$ stands for the realization number  of the random disorder)
broadens with increasing disorder strength $a$,
as is clearly seen in Fig.~\ref{Fig3}a (see also Appendix~\ref{SubError}). Physically,
this would result in an
 inhomogeneous broadening of the transmission spectrum
of a hypothetical large-area microcavity with
randomly displaced inner interfaces, where the displacement changes gradually on some large-distance scale,
and assuming incoherent addition of the transmission of different parts of this large microcavity.
Such inhomogeneous broadening was observed, e.g., in high-quality factor III-V nitride microcavities\cite{Christmann2006}
and attributed to  homogeneous areas (at a local scale of $\sim 8~\mu$m), separated by fluctuations occuring on a short distance scale.
The realization of high Q-factor in such microcavities is likely to be limited by the structural disorder~\cite{Gacevic2013,Gacevic2018}.

From comparison with Fig.~\ref{Fig3}b we see that for a disorder parameter $a=0.1505$, the
inhomogeneous broadening exceeds significantly the homogeneous linewidth, and it
 will be shown in the next section that the inhomogeneous broadening is equal to the homogeneous one
for $a \approx 0.02$.

\begin{figure}[h!]
\centering{\includegraphics[width=7cm]{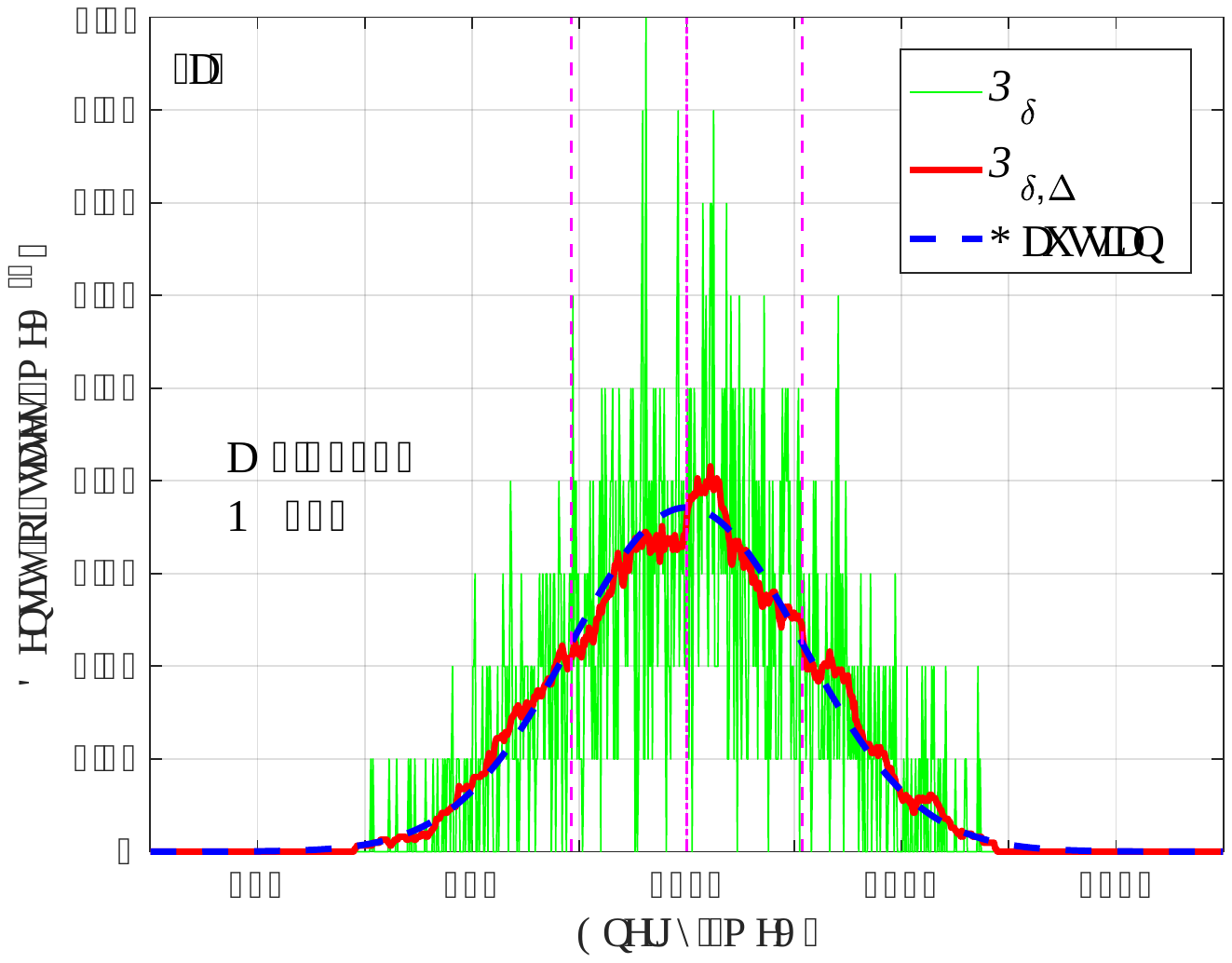}  \\
\includegraphics[width=7cm]{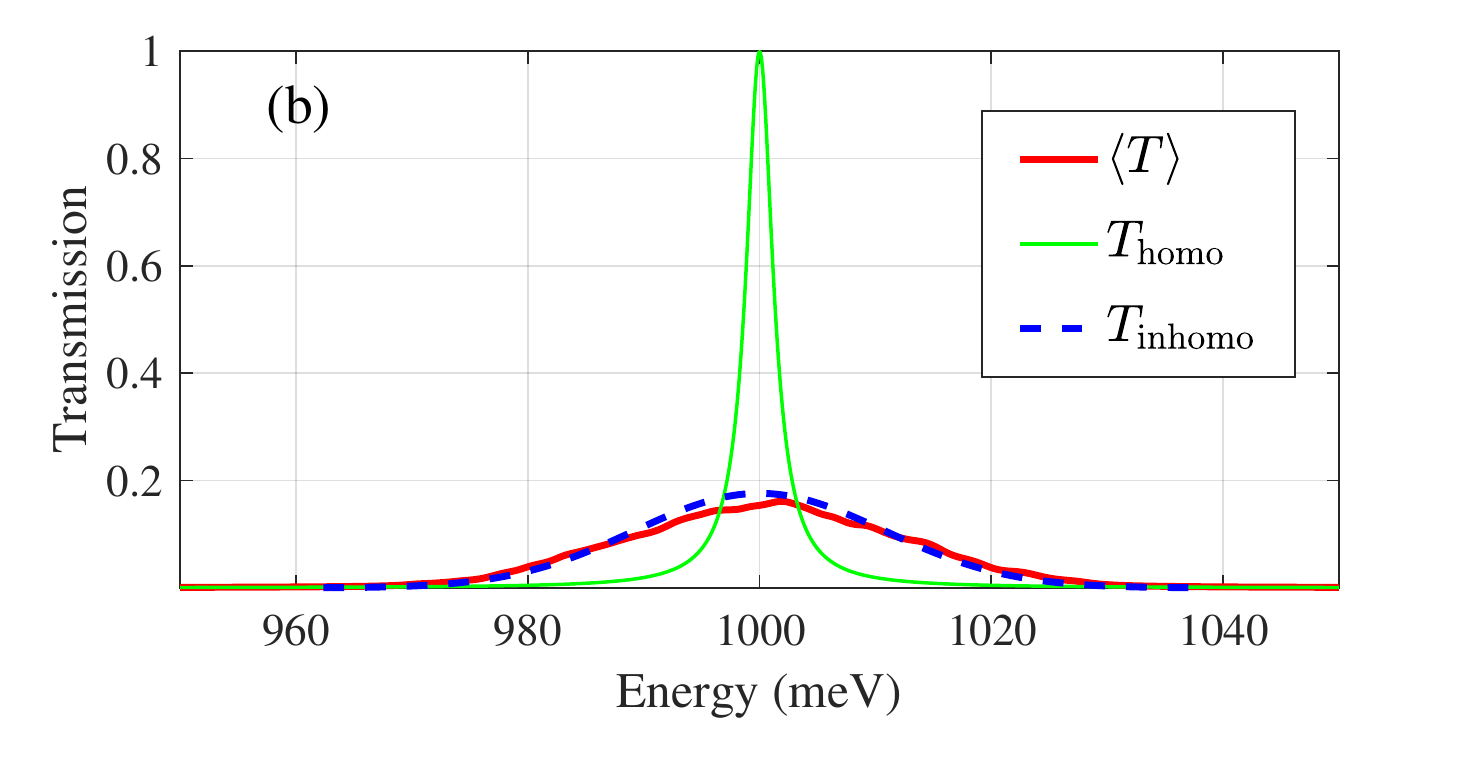}
\includegraphics[width=7cm]{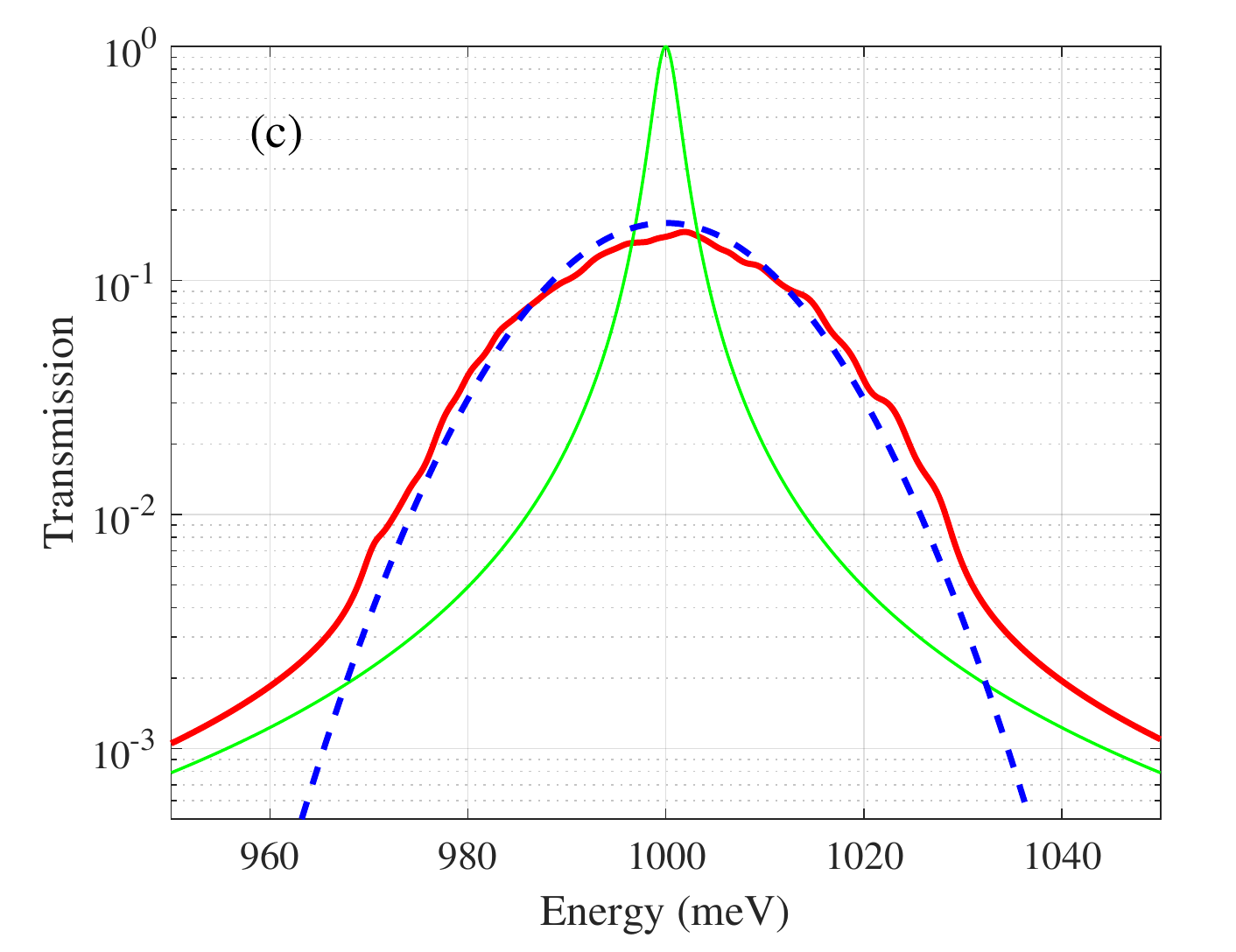}}
\caption{ \label{Fig4}
(Color online) (a) Densities of states [Eqs.~(\ref{Eq_PdE}) and (\ref{Eq_PdEav})], calculated with   $\delta = 0.1$~meV and $\Delta = 1.5$~meV
for the realizations of disorder in Fig.~\ref{Fig3}. Red vertical dashes denote
 $\Omega_0\pm \sigma_\Omega$ and the solid line $\Omega_0$; (b) Averaged inhomogeneously broadened
transmission spectrum of a large cavity with all realizations of disorder (red solid curve). The green solid line is the homogeneously broadened transmission spectra
of the ideal cavity calculated by Eq.~(\ref{T(E)}). Blue dashed curves display Gaussian distributions [Eqs. (\ref{Eq_Gaussian}) and (\ref{Tav})]; (c) same as panel (b) in logarithmic scale.
 }
\end{figure}

As to the distribution of resonance energies, as expected from the central limit theorem and the
superposition of 16 independent uniformly distributed random variables $\beta_j$ (see Eq.~1), it turns out to be close to
normal Gaussian.
A discretized density of states can be defined on an energy mesh  with a small
step $\delta < \Gamma_0$ as
\begin{eqnarray} \label{Eq_PdE}
P_\delta(E)=\frac{1}{\delta}\sum_{\nu} \int_{E-\delta/2}^{E+\delta/2} \delta(E'-\Omega_0(\nu)) dE',
\end{eqnarray}
where the sum is evaluated over all random realizations. It can be smoothed on a
larger energy scale by convolution with a normalized rectangular function of width $\Delta \approx \Gamma_0$, resulting in the distribution
\begin{eqnarray} \label{Eq_PdEav}
P_{\delta,\Delta}(E)=\frac{1}{\Delta}\int_{E-\Delta/2}^{E+\Delta/2} P_\delta(E') dE'.
\end{eqnarray}
Typical densities of states  Eqs.(\ref{Eq_PdE}) and (\ref{Eq_PdEav}) for the distribution of poles in Fig.~\ref{Fig3}a
with an amplitude of disorder of $a=0.1505$
are displayed as green and red lines in Fig.~\ref{Fig4}a for  $\delta = 0.1$~meV and $\Delta = 1.5$~meV.
The averaged density is very close
to the normal Gaussian  distribution
\begin{eqnarray} \label{Eq_Gaussian}
P_\mathrm{Gauss}(E)=\frac{1}{\sqrt{2\pi}\sigma_\Omega}\exp \left(-\frac{(E-\Omega_0^{(0)})^2}{2\sigma_\Omega^2}        \right),
\end{eqnarray}
plotted as a blue dashed line in Fig.~\ref{Fig4}a, in accordance with the Central Limit Theorem.
This theorem states that if you sum   up a large number of random variables, the distribution of the sum
will be approximately normal (i.e., Gaussian) under certain conditions, see, e.g., Ref.~\cite{Billingsley1995}.

The transmission spectrum of the ideal microcavity in the vicinity of the
fundamental cavity mode is approximated quite well by a Lorenzian
\begin{equation}\label{T(E)}
T(E,\Omega_0)=\frac{\Gamma_0^2}{(E-\Omega_0)^2+\Gamma_0^2} ,
\end{equation}
see  the solid and dashed lines in Fig.~\ref{Fig2}c. Thus, the inhomogeneously
broadened spectrum of a large microcavity with the distribution of resonances is expected to
exhibit the Voigt function~\cite{Voigt1912,Olver2010} shape,
\begin{equation}
  \langle T(E) \rangle  =  \int T(E,\Omega_0) P_\mathrm{Gauss}(\Omega_0) d \Omega_0
  .
 \end{equation}
 In the limit $\Gamma_0 \ll\sigma_\Omega $
 the averaged transmission is approximately Gaussian,
 \begin{equation}
 \label{Tav}
\langle T(E) \rangle \approx \frac{\Gamma_0}{\sigma_\Omega}\sqrt{\frac{\pi}{2}} \exp \left(-\frac{(E-\Omega_0^{(0)})^2}{2\sigma_\Omega^2}\right),
\end{equation}
except for the Lorenzian tails for $|E-\Omega_0| > \sigma_\Omega$.
An example of averaged spectra corresponding to the distribution of poles at
disorder parameter $a=0.1505$ for the convolution in Fig.~\ref{Fig4}a
is given in Fig.~\ref{Fig4}b and in logarithmic scale in Fig.~\ref{Fig4}c.
The averaged transmission spectra $\langle T(E)\rangle$ (red curves in panels b,c)
coincide quite well with the Gaussian spectra, Eq.~(\ref{Tav}) (blue dashed curves)
in the central part of the broadened resonance. In contrast, the Lorenzian tails approaching the homogeneously
broadened spectrum Eq.~(\ref{T(E)}) (solid green curves) are clearly
visible in panel c due to the log scale.

\begin{figure*}[h!]
\centering{\includegraphics[width=5.5cm]{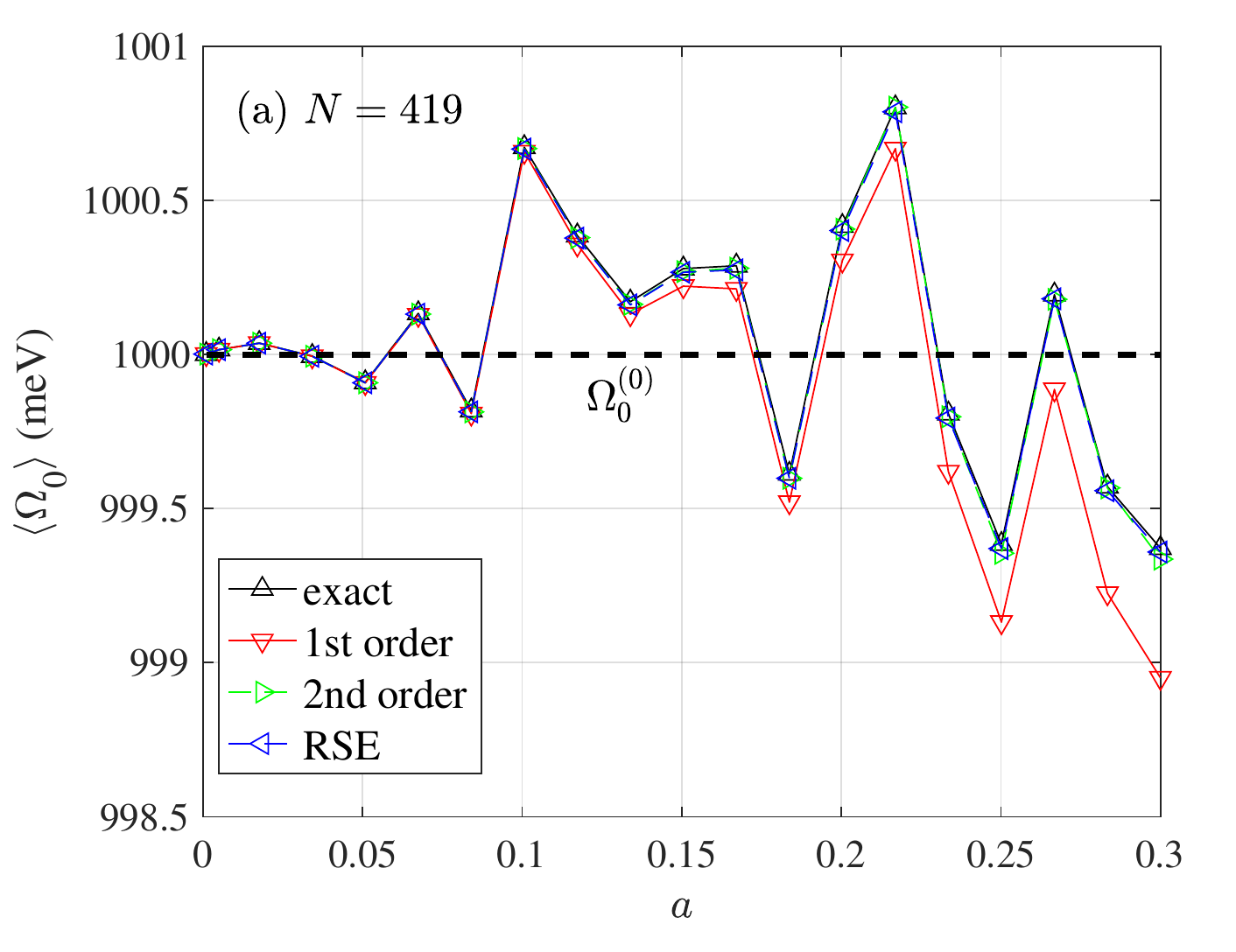}
\includegraphics[width=5.5cm]{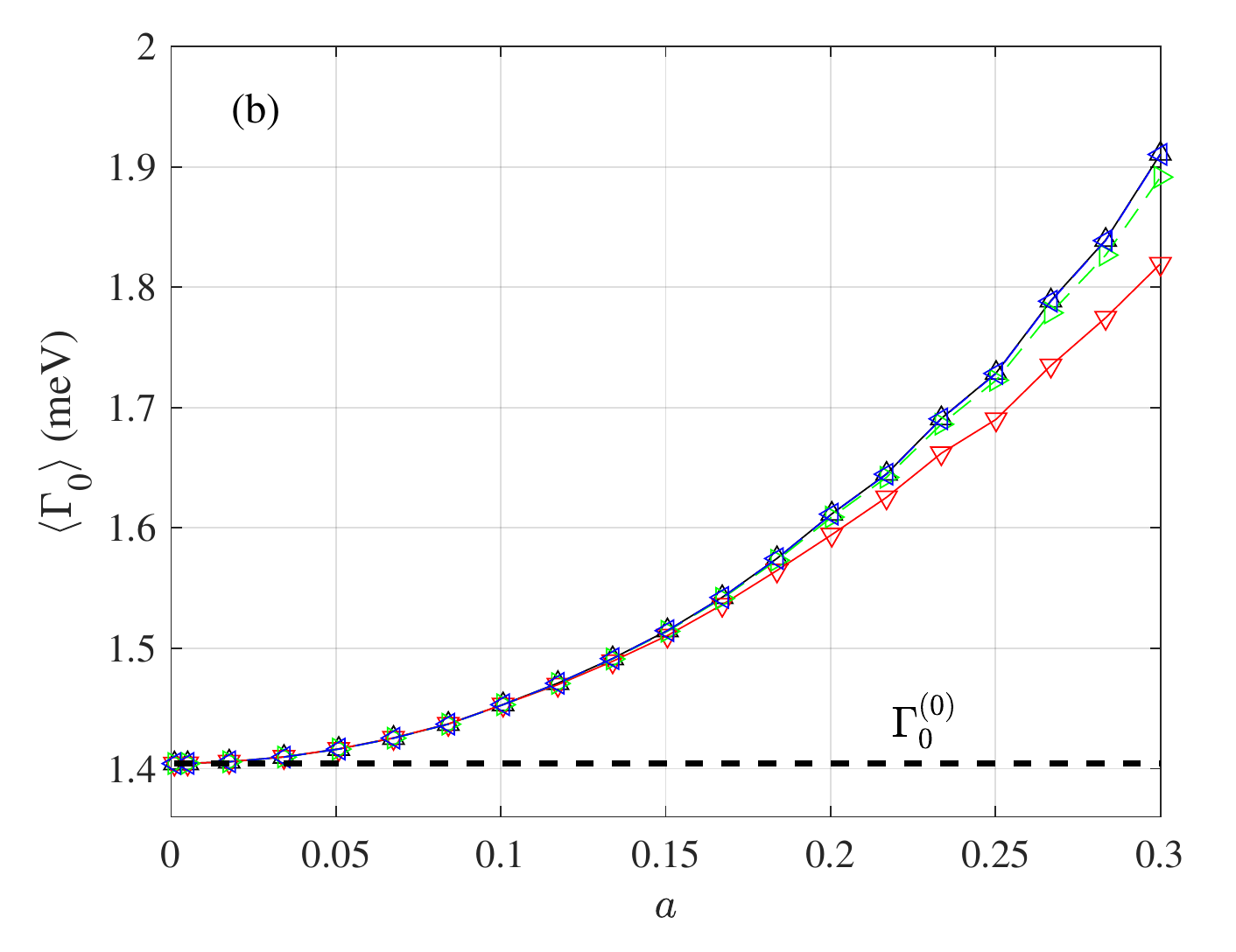}}
\centering{\includegraphics[width=5.5cm]{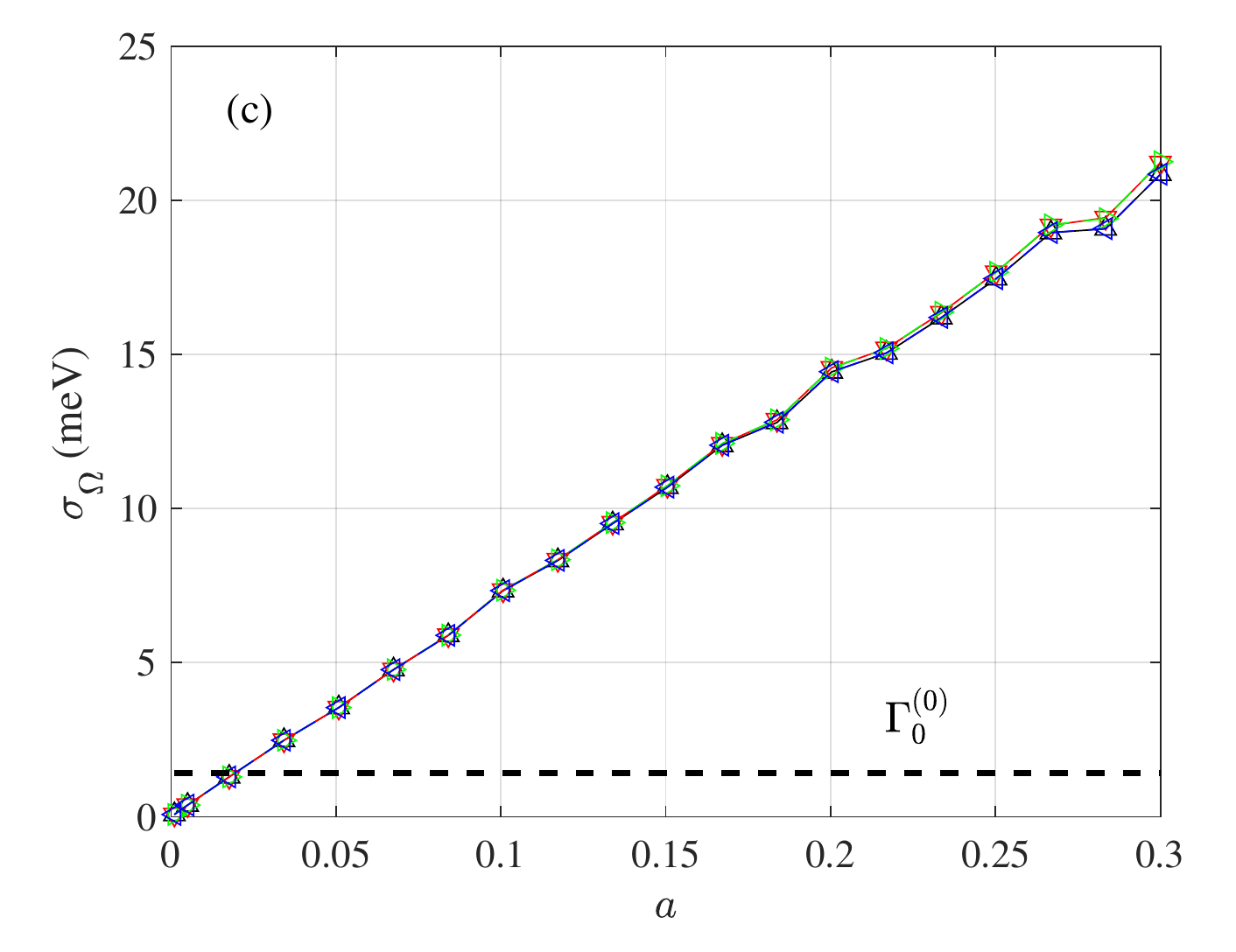}
\includegraphics[width=5.5cm]{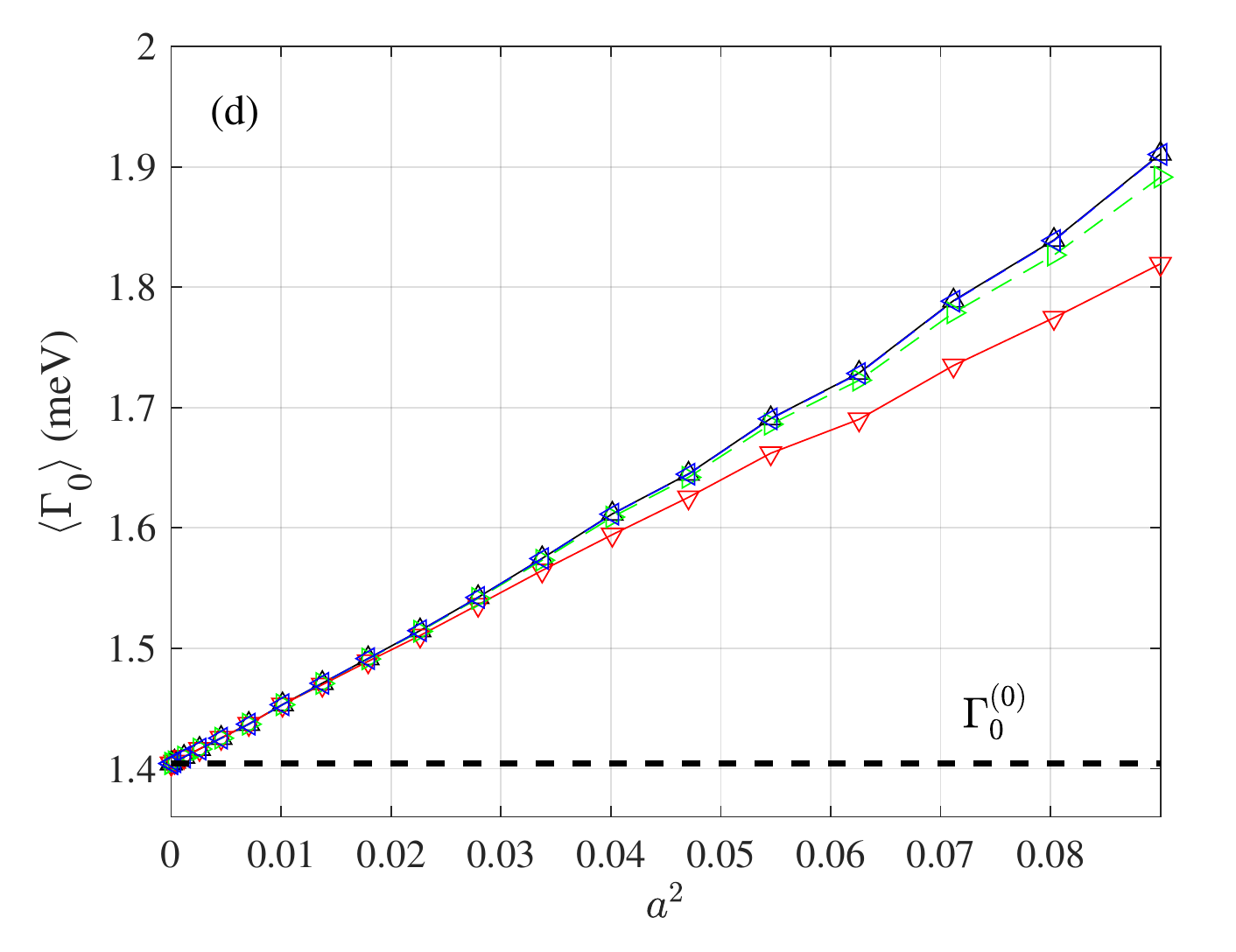}}
\caption{ \label{Fig5}
(Color online) Resonance energy $\langle \Omega_0\rangle$ (a) and half linewidth $\langle \Gamma_0\rangle$  (b) as functions of disorder parameter $a$, averaged over 1000 realizations of
random interface displacements. Panel (c) shows the standard deviation of resonance energy $\sigma_\Omega$ as a function of $a$, and panel (d) depicts
$\langle \Gamma_0\rangle$ as a function of $a^2$. The values of $\Omega_0^{(0)}$ [$\Gamma_0^{(0)}$] for the ideal microcavity without disorder are shown as horizontal
dashed  lines in panel (a) [panels (b-d)].
 }
\end{figure*}

\subsection{\label{SubVsA} Dependence on the disorder parameter}

The dependence of the averaged parameters of the fundamental cavity mode on the disorder parameter $a$ are illustrated in Fig.~\ref{Fig5}.
Panels a and b display the averaged  fundamental cavity mode energy  $\langle \Omega_0\rangle$
and half linewidth $\langle\Gamma_0\rangle$, respectively, as functions of the disorder parameter $a$.
The averaging is carried out over 1000 random realizations, \textit{different} for each value of $a$.
Panel c depicts the inhomogeneous broadening. It displays
the fundamental cavity mode energy standard deviation
$\sigma_\Omega=\langle \left(\Omega_0-\langle \Omega_0\rangle\right)^2\rangle^{1/2}$ as a
function of disorder parameter $a$. Panel d contains the same dependence as in panel b, but plotted instead as a function of $a^2$.

The averaged position of the resonance does not shift significantly with the growth of the disorder parameter $a$.
Fluctuations are due to the finite number of realizations used.
The magnitude of the inhomogeneous broadening, which is given by $\sigma_\Omega$, grows linearly with $a$, and the averaged half linewidth $\langle\Gamma_0\rangle$ grows quadratically with $a$
(the latter is clearly visible in panel d).
The inhomogeneous broadening matches the homogeneous linewidth of the resonance at $a\approx 0.02$.

\begin{figure}[h!]
\centering\includegraphics[width=6cm]{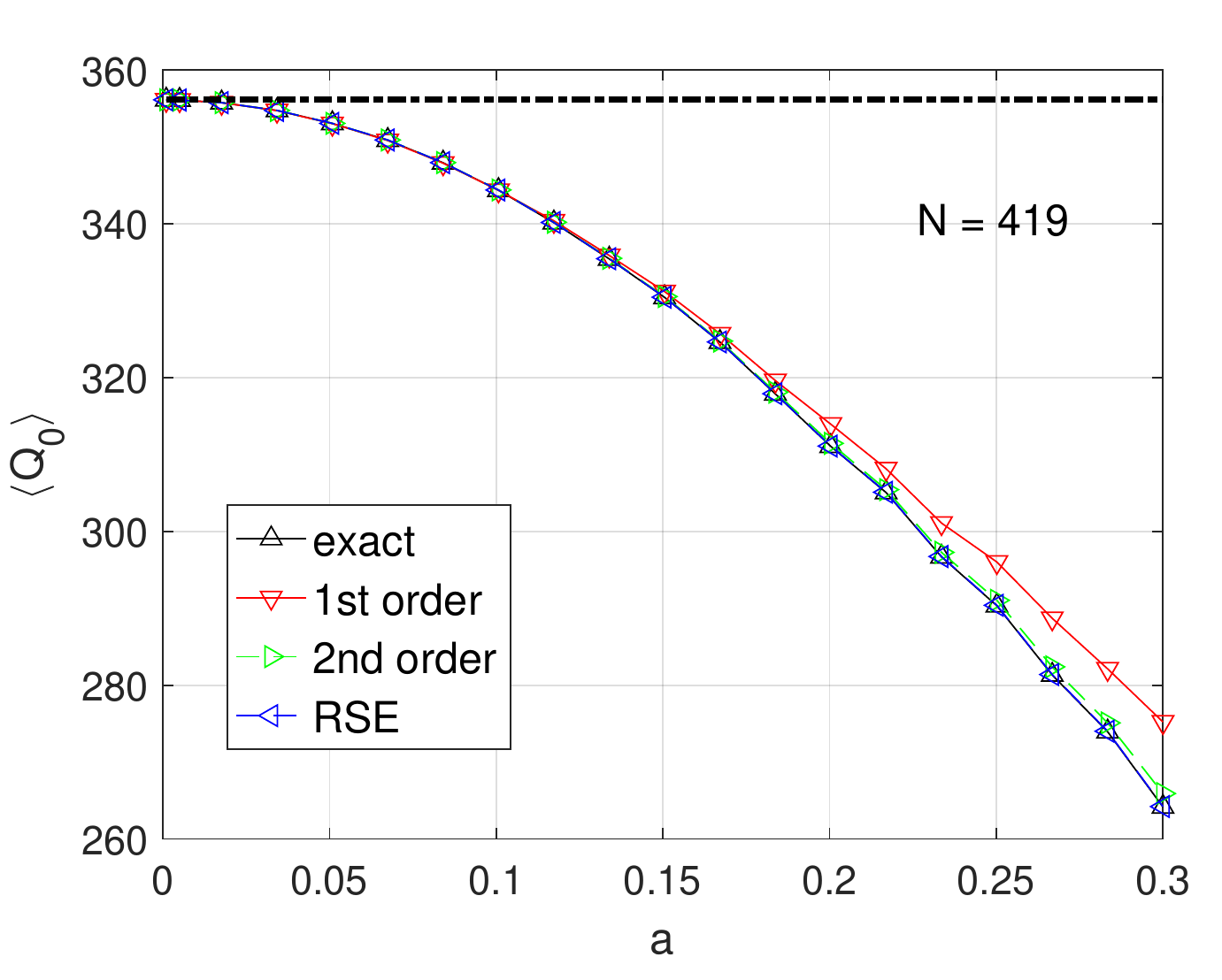}
\caption{ \label{Fig6}
(Color online) Dependence of the microcavity quality factor on the disorder parameter $a$, 
calculated exactly and in the first and second orders of perturbation theory,
as well as by the resonant-state expansion with 419 states.
The averaging is carried out over 1000 realizations of random displacements of inner interfaces in the microcavity.
 }
\end{figure}

The increase of the homogeneous linewidth results in a decrease of the averaged microcavity quality factor that
depends quadratically on the
disorder parameter $a$, as illustrated in Fig.~\ref{Fig6}.

\section{\label{Sec5} Discussion}

The reasons for the power  scaling $a^\alpha$ of
$\langle \Omega_0 \rangle$, $\sigma_\Omega$ and $\langle \Delta \Gamma_0 \rangle$
 with $\alpha = 0, 1$ and 2, respectively, can be
understood in the first-order approximation of the resonant-state expansion.

 The characteristic feature of the fundamental cavity mode electric field distribution for an
unperturbed microcavity with exactly $\lambda/4$ Bragg pairs,
exactly $\lambda$ cavity layer, and with the boundary conditions of Eq.~(\ref{Eq_rn})
can be clearly seen in Fig.~\ref{Fig1}.
Namely, the values of the real and imaginary parts of the electric eigenfield
 are subsequently zeroed exactly at successive interfaces. As a result,  in the vicinity of each interface, either the
 real or the imaginary part of the field is either a constant or a
linear function of the distance to this interface $z-z_{0,j}$ , i.e.,
\begin{eqnarray} \nonumber
\mathrm{Re}\, \mathcal{E}_0^{(0)}(z)& \approx & D_j + \mathcal{O}(z-z_{0,j}) , \\  \label{ReImEn1}
\mathrm{Im}\, \mathcal{E}_0^{(0)}(z)& \approx & (z-z_{0,j}) F_j  + \mathcal{O}\left((z-z_{0,j})^2\right),
\end{eqnarray}
or
\begin{eqnarray} \nonumber
\mathrm{Re}\, \mathcal{E}_0^{(0)}(z)& \approx & (z-z_{0,j}) D_j  + \mathcal{O}\left((z-z_{0,j})^2\right), \\  \label{ReImEn2}
\mathrm{Im}\, \mathcal{E}_0^{(0)}(z)& \approx & F_j + \mathcal{O}(z-z_{0,j}) ,
\end{eqnarray}
where $D_j, F_j$ are constants. The signs of $D_j, F_j$ are identical (negative or positive) on the
right-hand sides of the layers with larger dielectric susceptibility (i.e., for odd $j=2m+1$),
and opposite on the their left-hand sides (for even $j =2m$).
Note that $\Delta \varepsilon(z) = |\Delta\varepsilon| \mathrm{sign}(z-z_{0,j})$ on such right-hand side interfaces,
and  $\Delta \varepsilon (z) = - |\Delta\varepsilon| \mathrm{sign}(z-z_{0,j})$ on the left-hand side ones.
Additionally, the normalization constant of the fundamental cavity mode is real, as discussed in Appendix~\ref{AppA}.
All this results in the following equation for the fundamental cavity mode eigenenergy, averaged over random
realizations:
\begin{equation} \label{Om}
\langle E_0 \rangle = E_0 \left( 1-\frac{1}{2} \langle V_{00} \rangle \right)
 \equiv E_0 +  \overline{\Delta E_0},
\end{equation}
with
\begin{equation} \label{deltahOm}
\overline{\Delta E_0} = -\frac{E_0}{2} \sum_j \langle V_{00,j} \rangle ,
\end{equation}
where the sum is over all inner interfaces and
\begin{eqnarray}  \nonumber
& V_{00,j}  = C_0^{-2} \int_{z_{0,j}}^{z_{0,j}+a_j} \Delta \varepsilon(z) \mathcal{E}_0^2(z)dz  \\
& \approx  |\Delta \varepsilon | C_0^{-2} \left\{ \begin{array}{cc}
\pm D_j^2 a_j^3/3 \mp  F_j^2 a_j +i |D_jF_j| a_j^2 , & j = 2m+1 \\
\mp D_j^2 a_j \pm  F_j^2 a_j^3/3 +i |D_jF_j| a_j^2 , & j = 2m
                             \end{array}
\right.
\end{eqnarray}
After averaging the odd powers of $a_j$ vanish, and, as a result, we obtain in the first resonant-state expansion order
and up to the second order in $a$
 \begin{equation}\label{DeltaVSa}
\langle\Delta \Omega_0\rangle = 0 , \,\, \langle\Delta \Gamma_0\rangle \propto a^2,
\end{equation}
in agreement with the numerical results in Fig.~\ref{Fig5}a,b.
As to the inhomogeneous broadening of $\Omega_0$, due to the terms linear in $a_j$,
$\sigma_\Omega^2$ is proportional to $a^2$ and thus $\sigma_\Omega \propto a$,
 in agreement with the numerical results in Fig.~\ref{Fig5}c.

This shows in particular the well known fact that the unperturbed planar Bragg microcavity is an
\emph{optimized} structure from the point of view of the maximum  Q
factor  (or minimum of homogeneous half linewidth $\Gamma_0$): Any change of its structure causes a decrease of $Q$
and increase of $\Gamma_0$. In fact, in the case the unperturbed structure would not correspond to a mimumum of $\Gamma_0$ versus layer
thicknesses, a linear dependence of $\Gamma_0$ with $a$ would be present.

\section{\label{Sec6} Conclusion}

To conclude, we have demonstrated that introducing random shifts of interfaces in a standard planar Bragg microcavity
causes a growth of the inhomogeneous broadening of the fundamental cavity mode, linear in the
disorder strength $a$, which quantifies the relative
change of the layer thicknesses. In contrast, the linewidth
increases proportionally to $a^2$, with an according decrease of the quality factor.
The inhomogeneous broadening starts to exceed the homogeneous one at a certain value of disorder parameter,
which is $a\approx 0.02$ for the considered microcavity. The first-order perturbation theory within the resonant-state expansion works
accurately up to a disorder strength of
$a\approx 0.1$, especially for calculating  the resonance energy. Furthermore, it allows to find a quantitative scaling
of the microcavity parameters with disorder strength.

%% Code for appendices and equation numbers
\appendix

%\section*{Appendix A: Sample}
%\setcounter{equation}{0}
%\renewcommand{\theequation}{A{\arabic{equation}}}

%\begin{equation} a+b=c.
%\end{equation}
\section{Appendix A: Poles of the scattering matrix via linearlization}\label{AppA}
\setcounter{equation}{0}
\renewcommand{\theequation}{A{\arabic{equation}}}

\begin{figure*}[h!]
\centering{\includegraphics[width=13cm]{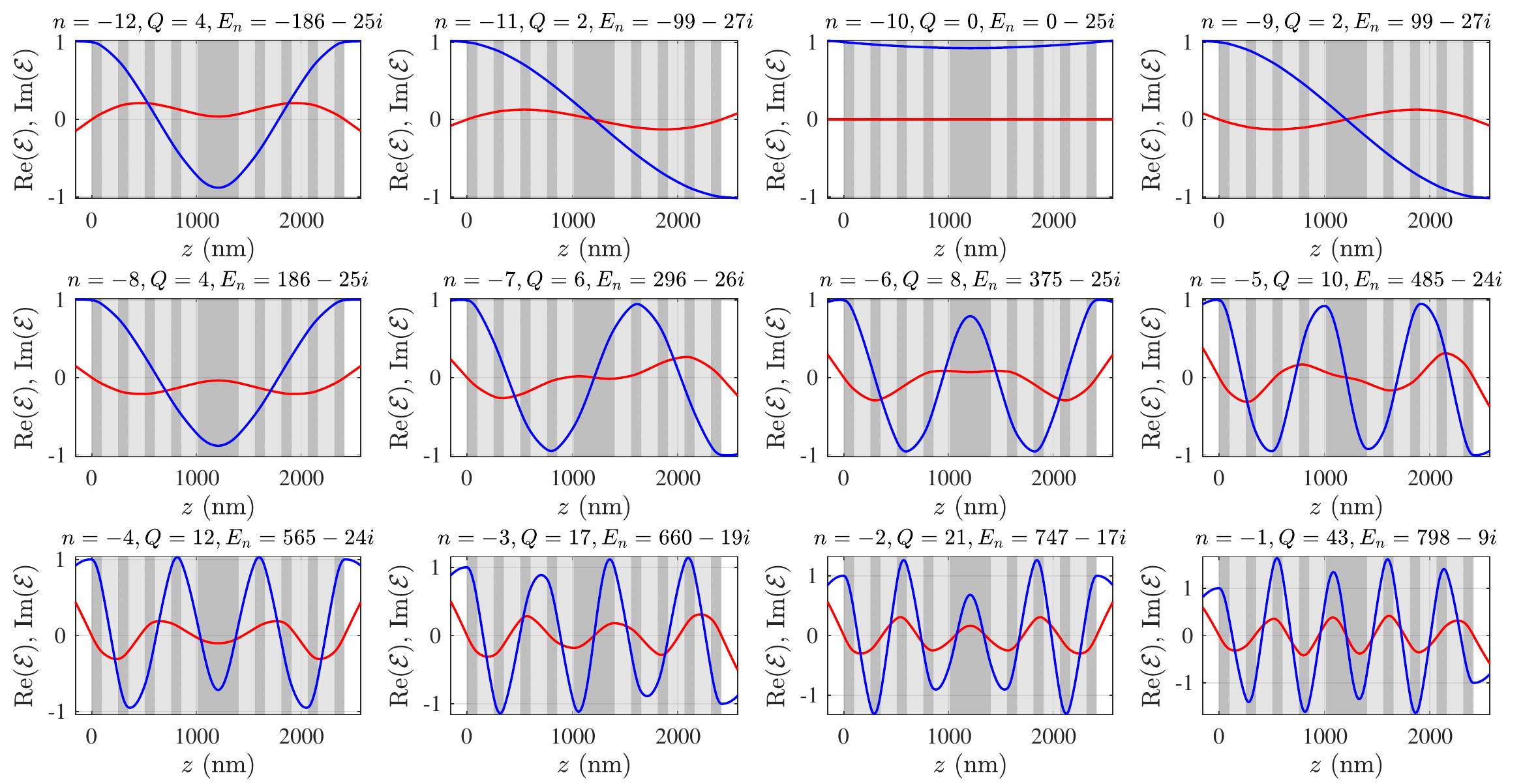}}
\includegraphics[width=13cm]{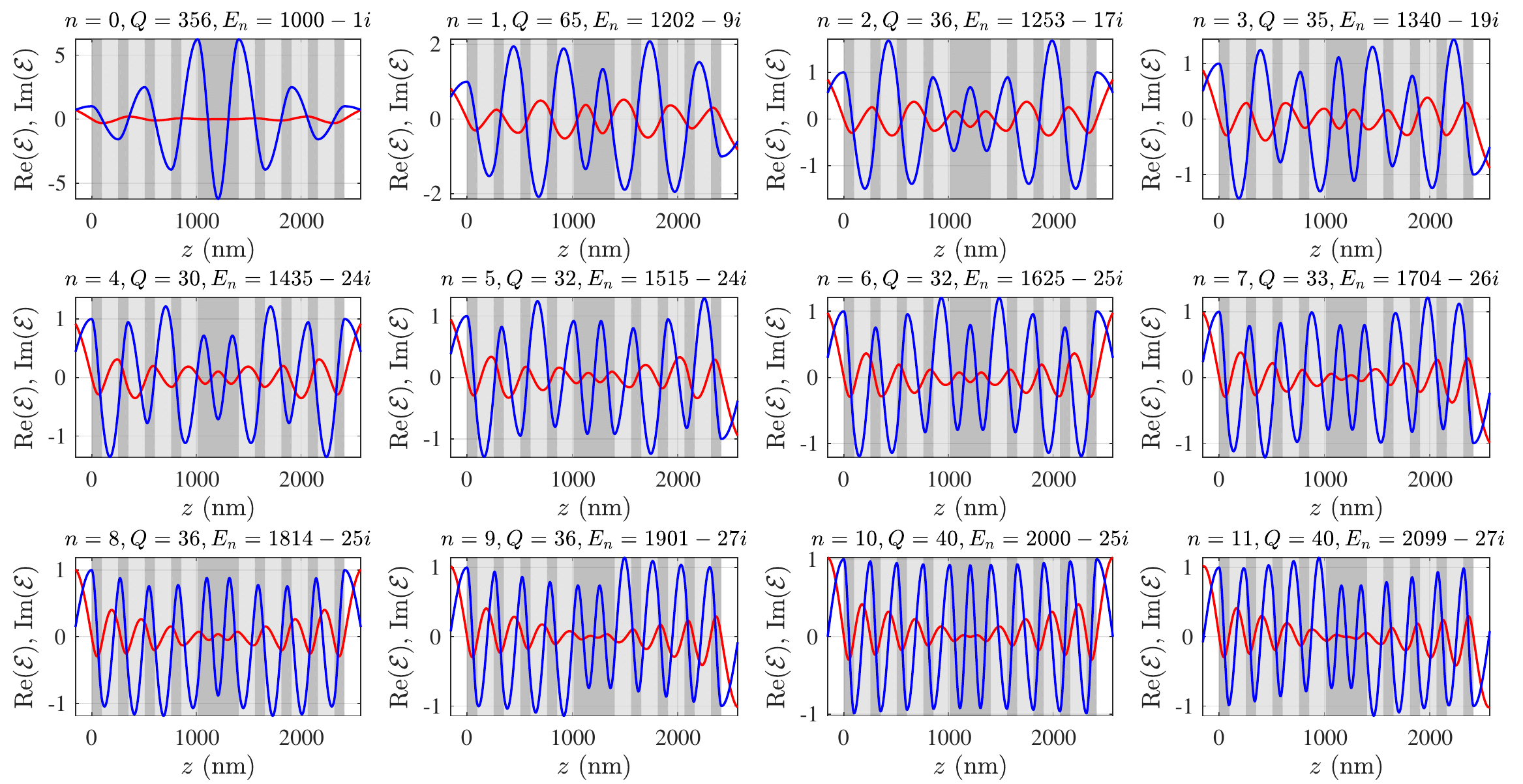}
\caption{ \label{Fig7}   %Fig7
(Color online) Real (blue solid lines) and imaginary (red solid lines) parts of the  electric field distributions
$\mathcal{E}_n^{(0)}(z)$  of the resonances of the ideal microcavity with $n=-12,-11,\ldots 11$, normalized according to conditions (\ref{Eq_rn}).
The Q-factors and eigenenergies $E_n  =  \Omega_n-i\Gamma_n$ [in milli-electron volt (meV)] are shown in the title of each panel.
 }
\end{figure*}

For normal incidence, the solutions of Maxwell equations for electric $\mathbf{E}$ and magnetic $\mathbf{H}$ fields in each layer
of the microcavity
\begin{equation}\label{Eq_EH}
\mathbf{E}=\left( E_x,0,0 \right), \mathbf{H}=\left(0,H_y,0\right),
\end{equation}
with
\begin{eqnarray}\label{ExHy}
E_x&=&A^+\exp(-i\omega t+ik_lz)+A^-\exp(-i\omega t-ik_lz), \\ \nonumber
H_y&=&n_lA^+\exp(-i\omega t+ik_lz)-n_lA^-\exp(-i\omega t-ik_lz),
\end{eqnarray}
where $n_l=\sqrt{\varepsilon_l}$ , $l=0,1,2$, $\varepsilon_0=1$ corresponds to semi-infinite
surrounding free space layers, and
$  k_l=n_l\omega/c$.
Note that we are using the Gaussian units. Defining the amplitude vector as
\begin{equation}\label{A}
|A\rangle=
    \left( \begin{array}{c}
           A^+ \\
           A^-
          \end{array}
    \right),
\end{equation}
the transfer matrix over a distance $d$
inside a homogeneous and isotropic  material is
\begin{equation}\label{Td}
  \tilde{T}_{l,d}= \left(\begin{array}{cc}
               \exp(ik_ld) & 0 \\
               0 & \exp(-ik_ld)
             \end{array}
             \right),
\end{equation}
with $|A(z+d)\rangle = \tilde{T}_{l,d}|A(z)\rangle$). The transfer matrix over the interface from material $l$ to $l'$  is
\begin{equation}\label{Td}
  T_{l',l}= \frac{1}{2}\left(\begin{array}{cc}
               1+K & 1-K \\

               1-K & 1+K
             \end{array}
             \right), \,\,\,
             K=\frac{n_l}{n_{l'}}\, .
\end{equation}

\begin{table}
\caption{\label{tab:tableA1}The eigenenergies $E_n = \Omega_n - i \Gamma_n$ and normalization constants
$C_n^2$ of 21 $-10 \leq n \leq 10$  resonances of the original ideal microcavity$^{a}$
}
%\begin{ruledtabular}
\centering\begin{tabular}{|c|c|c|c|c|}
$n$ & $\Omega_n$ (meV) & $\Gamma_n$ (meV) & $\mathrm{Re}(C_n^2) $ (nm) & $\mathrm{Im}(C_n^2)/\mathrm{Re}(C_n^2) $   \\
  \hline
-10 &  0      & 24.8 & 7.12$\cdot 10^{3}$ & 3.15$\cdot 10^{-17}$ \\
-9  &    99.2 & 26.5 & 6.69$\cdot 10^{3}$ & 3.17$\cdot 10^{-2}$ \\
-8  &   186.3 & 25.0 & 7.08$\cdot 10^{3}$ & 6.54$\cdot 10^{-2}$ \\
-7  &   295.6 & 25.7 & 6.88$\cdot 10^{3}$ & 9.99$\cdot 10^{-2}$ \\
-6  &   375.3 & 25.0 & 7.07$\cdot 10^{3}$ & 1.39$\cdot 10^{-1}$ \\
-5  &   485.3 & 23.7 & 7.45$\cdot 10^{3}$ & 1.84$\cdot 10^{-1}$ \\
-4  &   565.4 & 23.6 & 7.44$\cdot 10^{3}$ & 2.39$\cdot 10^{-1}$ \\
-3  &   659.5 & 19.1 & 9.20$\cdot 10^{3}$ & 2.98$\cdot 10^{-1}$ \\
-2 &   746.6 & 17.3 & 1.01$\cdot 10^{4}$ & 3.72$\cdot 10^{-1}$ \\
-1  &   797.9 & 9.18 & 1.90$\cdot 10^{4}$ & 4.28$\cdot 10^{-1}$ \\ \hline
0 &  1000.0 & 1.40 & 1.40$\cdot 10^{5}$ & 1.17$\cdot 10^{-9}$ \\ \hline
1 &  1202.0 & 9.18 & 1.90$\cdot 10^{4}$ &-4.28$\cdot 10^{-1}$ \\
2 &  1253.3 & 17.3 & 1.01$\cdot 10^{4}$ &-3.72$\cdot 10^{-1}$ \\
3 &  1340.4 & 19.1 & 9.20$\cdot 10^{3}$ &-2.98$\cdot 10^{-1}$ \\
4 &  1434.5 & 23.6 & 7.44$\cdot 10^{3}$ &-2.39$\cdot 10^{-1}$ \\
5 &  1514.6 & 23.7 & 7.45$\cdot 10^{3}$ &-1.84$\cdot 10^{-1}$ \\
6 &  1624.6 & 25.0 & 7.07$\cdot 10^{3}$ &-1.39$\cdot 10^{-1}$ \\
7 &  1704.3 & 25.7 & 6.88$\cdot 10^{3}$ &-9.99$\cdot 10^{-2}$ \\
8 &  1813.6 & 25.0 & 7.08$\cdot 10^{3}$ &-6.54$\cdot 10^{-2}$ \\
9 &  1900.7 & 26.5 & 6.69$\cdot 10^{3}$ &-3.17$\cdot 10^{-2}$ \\
10 &  2000.0 & 24.8 & 7.12$\cdot 10^{3}$ & 2.10$\cdot 10^{-8}$ \\
 \hline
 \multicolumn{5}{l}{$^{a}$\footnotesize{The parameters for the fundamental cavity mode with $n=0$ are indicated by a frame.}} \\
\end{tabular}
%\end{ruledtabular}

\end{table}

We can calculate the transfer matrix over the entire microcavity as
$$
T(\omega) = T_{0,1} \left(T_{BP}^{-1}\right)^4 \tilde{T}_{1,L_C}\left(T_{BP}\right)^4T_{1,0},
$$
where
$$
T_{BP}=T_{1,2}\tilde{T}_{2,L_2}T_{2,1}\tilde{T}_{1,L_1},
$$
so that the amplitude vectors from the left and right sides of the microcavity are connected as
\begin{equation}\label{psiT}
|A_L\rangle = \left(
\begin{array}{c}
 A_L^+\\ A_L^-
\end{array}\right)  ,\,\,\,
|A_R\rangle = \left(
\begin{array}{c}
 A_R^+\\ A_R^-
\end{array}\right), \,\,\,
|A_L\rangle = T(\omega)|A_R\rangle.
\end{equation}

\begin{figure}[h!]
\centering{\includegraphics[width=5.5cm]{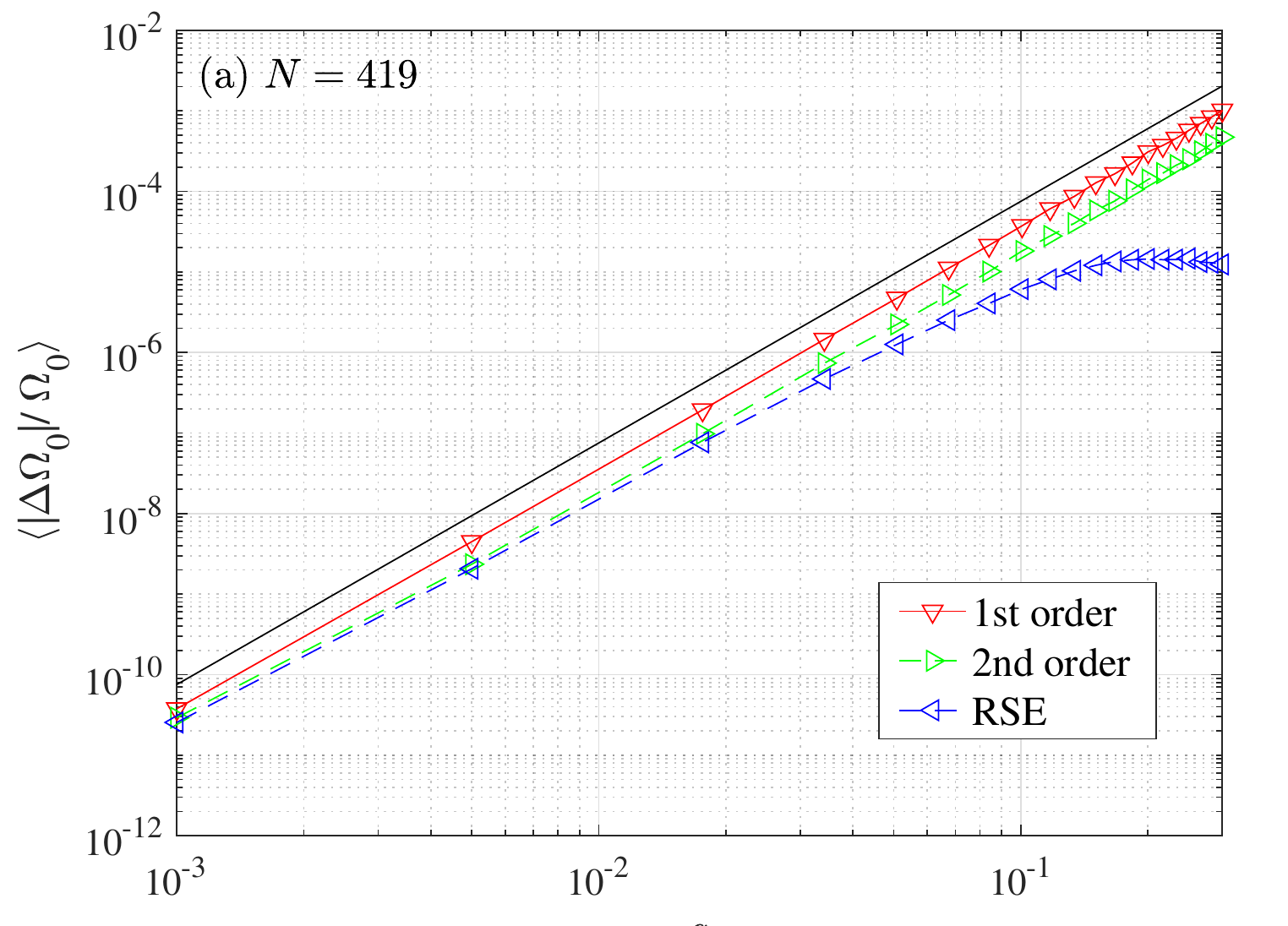}
\includegraphics[width=5.5cm]{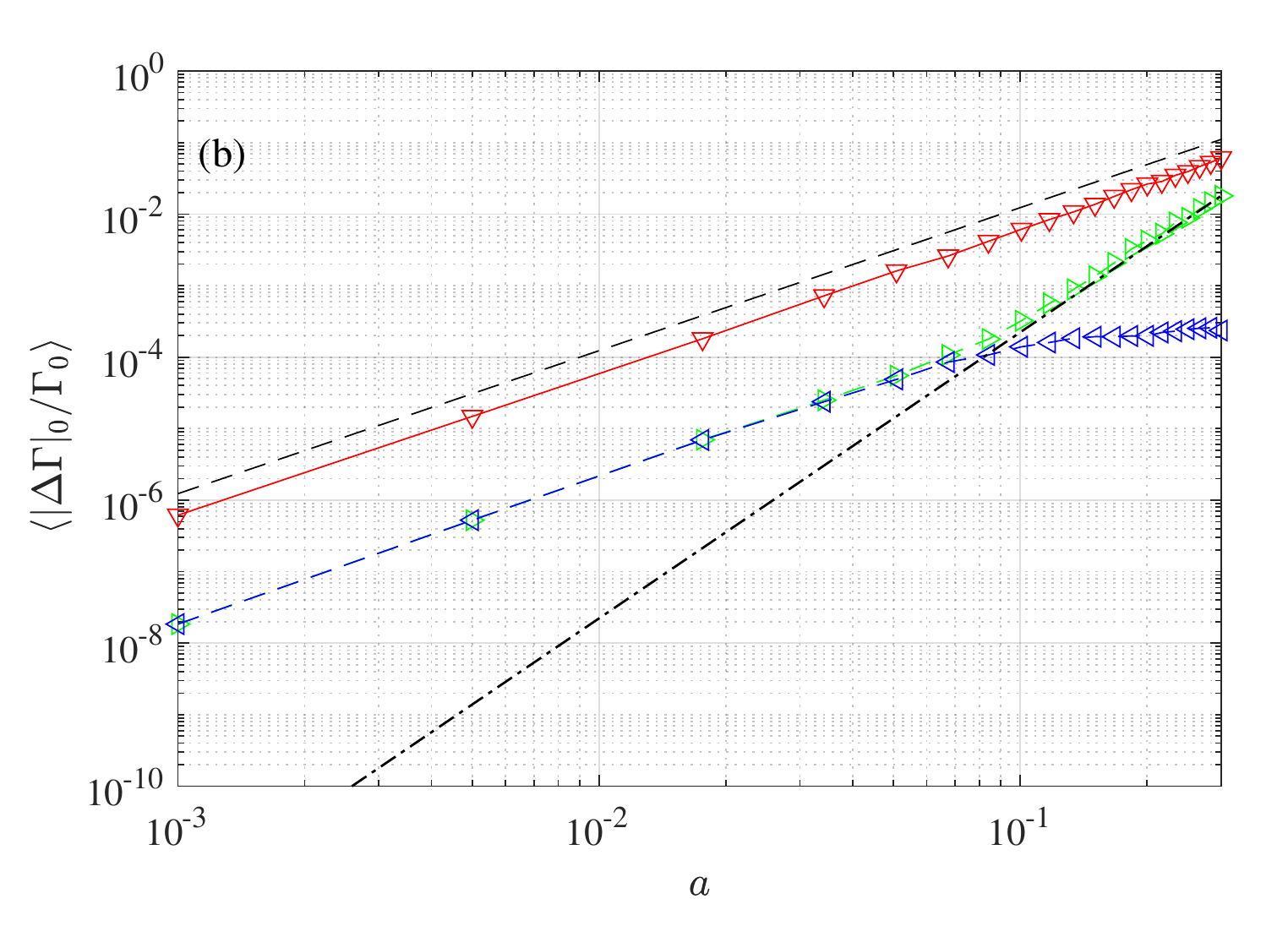}}
\caption{ \label{Fig8}
(Color online) Relative accuracy
of the first- and second-order perturbation theory, as well as the resonant-state expansion with 419 states
for real (a) and imaginary (b) parts of the
resonance energy as functions of disorder parameter $a$ for the fundamental cavity mode, averaged over 1000 realizations of
random interface displacements. Black dashed, solid, and dashed-dotted lines show $a^2$, $a^3$, and $a^4$  dependencies, respectively. The relative accuracy of
the quality factor is same as shown in panel (b). The relative accuracy is calculated as the relative difference between
the exact and the approximate methods.
 }
\end{figure}

\begin{figure}[h]
\centering{\includegraphics[width=5.5cm]{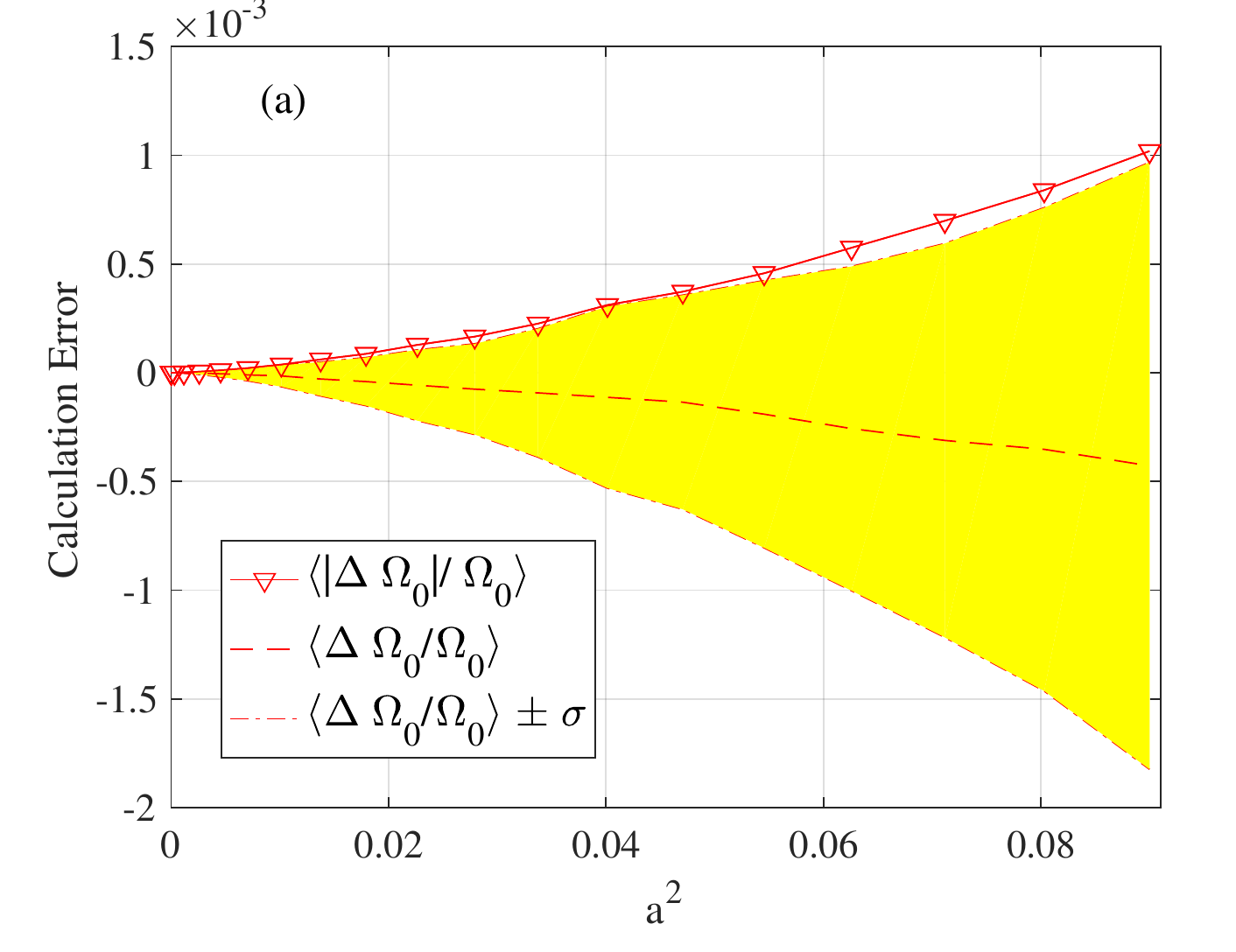}
\includegraphics[width=5.5cm]{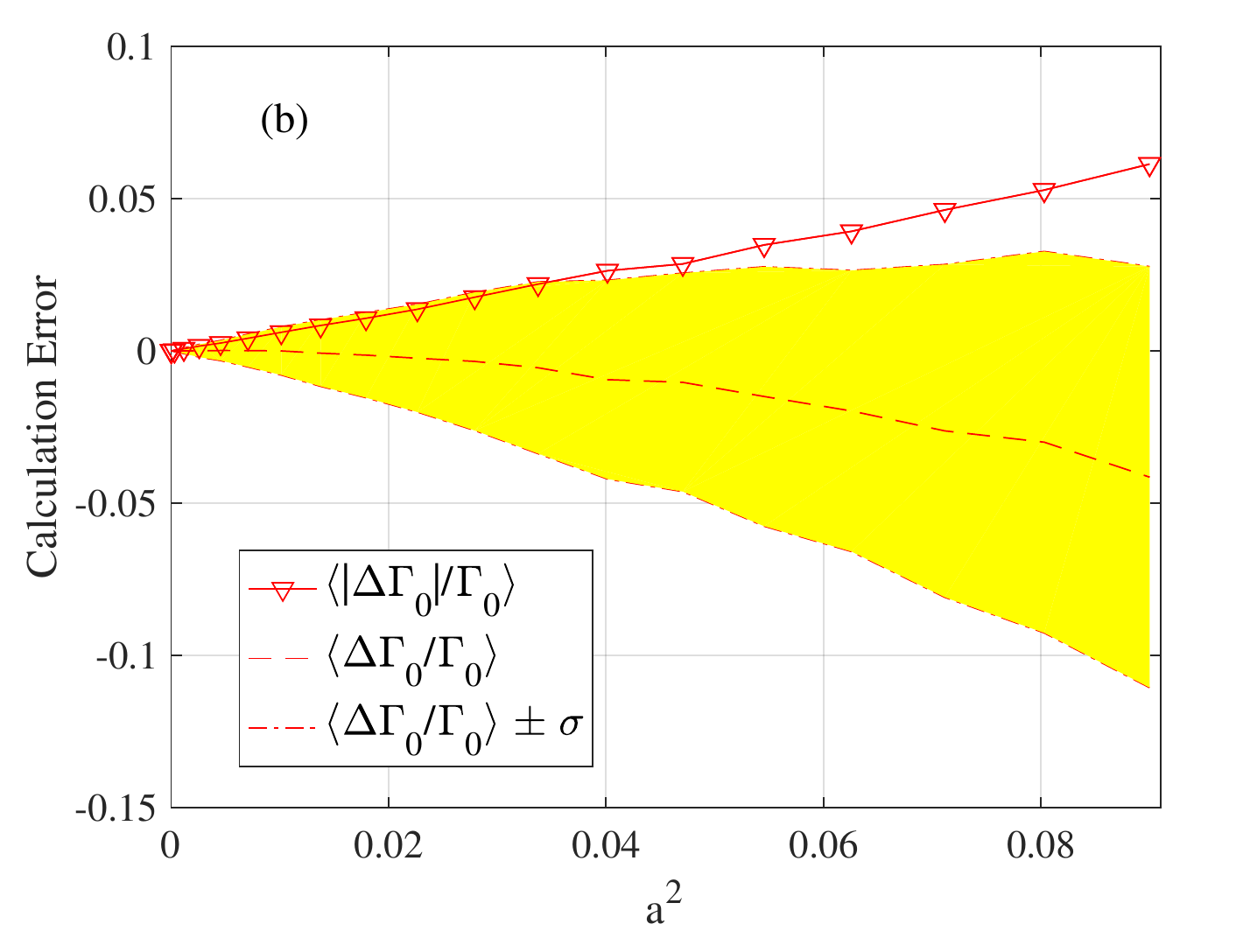}}
\caption{ \label{Fig9}
(Color online) Relative calculation error of the first-order  perturbation theory
for real (a) and imaginary (b) parts of the
resonance energy, as functions of $a^2$ for the fundamental cavity mode, averaged over 1000 realizations of
random interface displacements. The dashed lines with open triangles
are the same as shown in Fig.~\ref{Fig8}. The dashed red lines are averaged calculation errors $\langle\Delta \Omega_0/\Omega_0\rangle$
and $\langle\Delta \Gamma_0/\Gamma_0\rangle$. Yellow regions show the width of the error distribution, e. g.,
 $\langle\Delta \Gamma_0/\Gamma_0\rangle \pm \sigma_{\Delta}$ in panel b.
 }
\end{figure}

Using the vectors of incoming and outgoing amplitudes
\begin{equation}\label{psiS}
|\mathrm{in}\rangle = \left(
\begin{array}{c}
 A_L^+\\ A_R^-
\end{array}\right)  ,\,\,\,
|\mathrm{out}\rangle  = \left(
\begin{array}{c}
 A_L^-\\ A_R^+
\end{array}\right) \,
\end{equation}
the optical scattering matrix is defined as
\begin{equation}\label{S-matrix}
|\mathrm{out}\rangle = S(\omega)
|\mathrm{in}\rangle .
\end{equation}
From this definition, it is seen that the physical meaning of the scattering matrix components
is
\begin{equation}\label{Srt}
 S=\left(
\begin{array}{cc}
r_{LL} & t_{RL} \\
t_{LR} & r_{RR}
\end{array}
\right),
\end{equation}
where, e.g., $r_{LL}$ is the amplitude reflection coefficient from the left side of microcavity to left,
and $t_{RL}$ is the amplitude transmission coefficient from left to right.
The connection with the components of transfer matrix is
\begin{equation}\label{StoT}
S=\left(
\begin{array}{cc}
-T_{22} ^{-1}T_{21} & T_{22} ^{-1} \\  T_{11}-T_{12}T_{22}^{-1}T_{21}  & T_{12}T_{22} ^{-1}
\end{array}
\right),
T= \left(
\begin{array}{cc}
T_{11} & T_{12} \\ T_{21} & T_{22}
\end{array}
\right)
.
\end{equation}
Eigensolutions (resonances) are found as
nonvanishing outgoing solutions $|\mathrm{out}\rangle = |\mathrm{o}_n\rangle \neq 0 $ at zero
input $|\mathrm{in}\rangle = 0$, which results in
the homogeneous equation for the resonant outgoing eigenvectors $|\mathrm{o_n}\rangle$ and eigenfrequencies $\omega_n$:
\begin{equation}\label{R-matrix}
S^{-1}(\omega_n)|\mathrm{o_n}\rangle \equiv R(\omega_n)|\mathrm{o_n}\rangle = 0.
\end{equation}
Equation~(\ref{R-matrix}) can be solved iteratively via a frequency-dependent linearization,
described, e.g., in Ref.~\cite{Gippius2005}.
Assuming that
$$\omega_n = \omega + \Delta\omega, $$
and linearizing Eq.~(\ref{R-matrix}) over $\Delta\omega$, we obtain
$$ 0 = R(\omega_n)\, |\mathrm{o}_n\rangle = R(\omega)\, |\mathrm{o}_n\rangle + \Delta\omega\frac{dR(\omega)}{d\omega}|\mathrm{o}_n\rangle , $$
which requires
$$R(\omega)\, |\mathrm{o}_n\rangle =- \Delta \omega  \frac{dR(\omega)}{d\omega}|\mathrm{o}_n\rangle .$$
Thus, we arrive at a linear $2{\times}2$ matrix problem to find $\Delta \omega$:
\begin{equation}\label{Delta}
W\, |\mathrm{o}_n\rangle =\Delta\omega  |\mathrm{o}_n\rangle,
\end{equation}
where the matrix
\begin{equation}\label{W}
W(\omega) \equiv - \left[ \frac{dR(\omega)}{d\omega}\right] ^{-1}R(\omega) = S(\omega) \left[ \frac{dS(\omega)}{d\omega}\right] ^{-1}
\end{equation}
can be easily calculated and  diagonalized. The latter equation follows from
$ d \left(S S^{-1}\right)/d \omega =0 $. The minimum eigenvalue $\Delta\omega$ of $W$
generates the corrected frequency $\omega'=\omega + \Delta\omega$, which is presumably closer to the solution
of Eq.~(\ref{S-matrix}). The procedure can be iteratively
continued until finding the solution with the desired accuracy. As a starting point
for iterations, it makes sense to use the real values of frequency, that correspond to the transmission
maxima (see in Fig.~\ref{Fig2}).

As for the resonance eigenvector, it is known in the case of
 mirror-symmetric structure in advance  due to symmetry constraints:
\begin{equation}\label{o_r}
 |\mathrm{o}_n\rangle = \left(
 \begin{array}{c}
 1\\ (-1)^{p_n}
\end{array}\right)    .
\end{equation}
The parity is $p_n=0$ for even and 1 for odd eigenfunctions.
The resonance distribution of the electric field can be then reconstructed easily as
\begin{equation}\label{Efield}
\mathcal{E}_n(z) = A_n^+(z)\exp(ik_nz)+ A_n^-(z)\exp(-ik_nz),
\end{equation}
with
\begin{equation}\label{Ar}
 \left(
 \begin{array}{c}
 A_n^+(z)\\ A_n^-(z)
\end{array}\right) = T_z(\omega_n)
\left(
 \begin{array}{c}
 0\\ 1
\end{array}\right),
\end{equation}
where $T_z(\omega_n)$ is the transfer matrix from the left side of the microcavity to point $z$ inside.

The calculated eigenenergies and normalization constants for the 21 resonances around the resonance with $\Omega_0=1$eV  for our microcavity are
given in Tab.~\ref{tab:tableA1}. The resonance at $\Omega_0=1$eV has the maximal quality factor.
In the main text we call it the fundamental cavity mode.
The electric eigenfields for $-12 \leq n \leq 11$
are shown in Fig.~\ref{Fig7}.
The resonance with $n=-10$ is `static', $\Omega_{-10}=0$. All other resonances are mirror-symmetric on the complex energy plane around it:
the resonances with  $\tilde{n}=n+10<0$ are $\Omega_{\tilde{n}-10}=-\Omega_{-\tilde{n}-10}<0$,  $\Gamma_{\tilde{n}-10}=\Gamma_{-\tilde{n}-10}$, and the eigenfields
are complex conjugate, i.e., $\mathrm{Re}\,\mathcal{E}_{\tilde{n}-10}(z)=\mathrm{Re}\,\mathcal{E}_{-\tilde{n}-10}(z)$, $\mathrm{Im}\,\mathcal{E}_{\tilde{n}-10}(z)=-\mathrm{Im}\,\mathcal{E}_{-\tilde{n}-10}(z)$.
The parity of the resonance with odd (even) $n$ is odd (even). The latter is the consequence
 of the mirror symmetry of the microcavity and the definition of normalized resonant states
using the boundary conditions Eq.~\ref{o_r}. Note that the normalization constants $C_n$ are, generally, complex (except those of the fundamental cavity mode and other
high-Q states, see below).
We use in the main text up to $N=419$ states
in the resonant-state expansion basis, positioned symmetrically around the fundamental cavity mode, i.e., with $\Omega_n$ for $-(N-1)/2 \leqslant n \leqslant (N-1)/2 $.

An interesting point about the normalization constant of the fundamental cavity mode is that it appears to be
real within the accuracy of our numerical calculation. In fact,
$$
C_{0,1}= \int_0^L \varepsilon (x) \mathcal{E}_{0}^2(x) dx \approx 1.4004{\cdot}10^{5} - 1.9733{\cdot}10^{2}i
$$
and
$$
C_{0,2}= \frac{i}{2k_0}\left[\mathcal{E}_{0}^2(0)+\mathcal{E}_{0}^2(L) \right] \approx -2.7708{\cdot}10^{-1} + 1.9733{\cdot}10^{2}i
$$
for the eigenfield $\mathcal{E}_{0}$, shown in Fig.~\ref{Fig1}. This field is normalized according to
$$
\mathcal{E}_0^{(0)}(0) = \mathcal{E}_0^{(0)}(L)=1,
$$
which follows from Eq.~(\ref{o_r}).
It appears that for the fundamental cavity mode with $n=0$
$$
C_0^2=C_{0,1}+C_{0,2} \approx 1.4004{\cdot}10^{5} + 1.6502{\cdot}10^{-4}i,
$$
so that $C_0$ is real with the accuracy of our numerical procedure.

%\section*{Appendix B: Sample}
%\setcounter{equation}{0}
%\renewcommand{\theequation}{B{\arabic{equation}}} %change B as needed
%\begin{equation}
%x-y=z.
%\end{equation}

\section{\label{SubError} Appendix B: Accuracy of  different approximations of the resonant-state expansion}

The averaged absolute values of relative errors for calculating $\Omega_0$ and $\Gamma_0$
by the first- and second-order approximations and the resonant-state expansion with 419 nearest resonant states are illustrated in Fig.~\ref{Fig8}
(panels a and b, respectively) as functions of the disorder parameter $a$.
It can be seen that the first-order perturbation theory becomes, as expected, less accurate with increasing disorder parameter, but it gives in most cases
quite accurate results, especially for the calculation of $\Omega_0$, and for small amplitude of disorder, $a<0.1$.

Figure~\ref{Fig8}a contains also, as a guide for the eye, black solid line, proportional to $a^3$, and Fig.~\ref{Fig8}b contains black dashed and dashed-dotted
ones, proportional to $a^2$ and $a^4$, respectively.
The magnitude of the calculation errors grows as $a^3$ and $a^2$ for the first perturbation order over the investigated range
 of $a$  for $\Omega_0$ and $\Gamma_0${, respectively.
For $\Omega_0$, the second order only provides a factor of 2 improvement and is limited by the basis size used in the resonant-state expansion.
The calculation error in the second perturbation order scales instead as $a^4$ for $\Gamma_0$, but is limited for small $a$ by the finite size of the resonant-state expansion basis
used and merges with the error of the full resonant-state expansion.
As to the calculation error of the full resonant-state expansion, in the case of $\Gamma_0$ it saturates around $2{\times}10^{-4}$
for  $a > 0.1$. For $a \lesssim 0.02$ the full resonant-state expansion error
coincides with that of the second order perturbation theory which means that the full resonant-state expansion becomes redundant.
However, the value of $a$ where the second order matches the full resonant-state expansion  depends on the basis size.
The saturated accuracy of the full resonant-state expansion for larger $a$
depends on the chosen basis size.
With decrease of the size of the resonant state basis this saturated accuracy worsens, e.g., to $\sim 2{\times}10^{-3}$
for $N=219$. Note that we take here the resonant state basis set symmetrical around the fundamental cavity mode.

As a result, the first-order perturbation theory of resonant-state expansion works well for $a<0.3$. Note that such a large $a$ corresponds to the
amplitude of interface displacement of up to 30\% of the thinner Bragg layer thickness, or in the present case
as large as $\sim 30$~nm. The averaged calculation error of the first-order perturbation
 is still smaller than 10\% for $a=0.3$. Of course, as can be understood from Fig.~\ref{Fig2} and  Fig.~\ref{Fig3}, there occur relatively rare
displacement realizations with a very large calculation error. However, the majority of disorder realizations is still reasonably well described by first-order  perturbation theory.
Figure~\ref{Fig9} illustrates the width of the range within which more than half of the disorder realizations are confined (filled by yellow color). With growing
disorder parameter systematic errors  arise $\langle\Delta \Omega_0/\Omega_0\rangle < 0$ and $\langle\Delta \Gamma_0/\Gamma_0\rangle < 0$. However, for weak disorder
these systematic errors are small, and $\langle | \Delta \Omega_0 |/\Omega_0\rangle \approx \sigma_{\Delta \Omega / \Omega}$,
$ \langle | \Delta \Gamma_0 |/\Gamma_0\rangle \approx \sigma_{\Delta \Gamma/\Gamma}$.

\section*{Funding} Deutsche Forschungsgemeinschaft (Mercator-Fellowship, SPP 1839); Russian Science Foundation (16-12-10538$\Pi$).

\section*{Acknowledgments}
The authors acknowledge support from DFG, and Russian Academy of Sciences.
S.G.T. and N.A.G. thank the Russian Science Foundation (Grant No. 16-12-10538$\Pi$) for support in part of the
calculations of the exact microcavity resonant states.

\section*{Disclosures}
The authors declare no conflicts of interest.

%%%%%%%%%% If using BibTeX:
%\bibliography{d:/localtexmf/bibtex/bib/photonics_main_A,d:/localtexmf/bibtex/bib/gen_phys_L}
%\end{document}

%%%%%%%%%% If preparing manually:
% \newcommand{\enquote}[1]{``#1''}

% \bibitem{Zhang:14}
% Y.~Zhang, S.~Qiao, L.~Sun, Q.~W. Shi, W.~Huang, L.~Li, and Z.~Yang,
%   \enquote{Photoinduced active terahertz metamaterials with nanostructured
%   vanadium dioxide film deposited by sol-gel method,}
%   {\protect\JournalTitle{Optics Express}} \textbf{22}, 11070--11078 (2014).

% {Optical Society}, \enquote{{OSA Publishing},}
%   \url{http://www.osapublishing.org}.

% \bibitem{FORSTER2007}
% P.~Forster, V.~Ramaswamy, P.~Artaxo, T.~Bernsten, R.~Betts, D.~Fahey,
%   J.~Haywood, J.~Lean, D.~Lowe, G.~Myhre, J.~Nganga, R.~Prinn, G.~Raga,
%   M.~Schulz, and R.~V. Dorland, \enquote{Changes in atmospheric consituents and
%   in radiative forcing,} in \enquote{Climate Change 2007: The Physical Science
%   Basis. Contribution of Working Group 1 to the Fourth assesment report of
%   Intergovernmental Panel on Climate Change,}  S.~Solomon, D.~Qin, M.~Manning,
%   Z.~Chen, M.~Marquis, K.~B. Averyt, M.~Tignor, and H.~L. Miler, eds.
%   (Cambridge University Press, 2007).

% \end{thebibliography}

\end{document}